\pdfoutput=1
\documentclass[a4paper,american,citeautoscript,floatfix,pdftex,aip,jcp,showpacs,superscriptaddress,%
longbibliography,%
reprint,twocolumn]{revtex4-1}

\usepackage{amsmath,amssymb}
\usepackage{graphicx}
\usepackage[T1]{fontenc}
\usepackage[utf8]{inputenc}
\usepackage{microtype}
\usepackage{textcomp}
\usepackage[textsize=footnotesize]{todonotes}
\usepackage{xspace}
\usepackage{hyperref, hypernat}
\usepackage{siunitx}
\usepackage[todos]{CMImacros}

\newlength{\figwidth}
\DeclareSIUnit \ppm{ppm}
\DeclareSIUnit[per-mode=symbol]\kvcm{\kilo\volt\per\centi\meter}
\DeclareSIUnit[per-mode=symbol]\wsqcm{\watt\per\square\centi\meter}

\newcommand{\pket}[1]{\ensuremath{\left|#1\right>_\textup{p}}\xspace}%
\newcommand{\tket}[2][]{\ensuremath{\left|#2\right>_{\textup{t}#1}}\xspace}%

\newcommand{\cfeldesy}{\affiliation{Center for Free-Electron Laser Science (CFEL), Deutsches
      Elektronen-Synchrotron DESY, Notkestrasse 85, 22607 Hamburg, Germany}}%
\newcommand{\uhhcui}{\affiliation{The Hamburg Center for Ultrafast Imaging, University of Hamburg,
      Luruper Chaussee 149, 22761 Hamburg, Germany}}%
\newcommand{\uhhphys}{\affiliation{Department of Physics, University of Hamburg, Luruper Chaussee
      149, 22761 Hamburg, Germany}}%
\newcommand{\granada}{\affiliation{Instituto Carlos I de F\'{\i}sica Te\'orica y Computacional and
      Departamento de F\'{\i}sica At\'omica, Molecular y Nuclear, Universidad de Granada, 18071
      Granada, Spain}}%

\begin{document}
\title{Adiabatic mixed-field orientation of ground-state-selected carbonyl sulfide molecules}%
\author{Jens S.\ Kienitz}\cfeldesy\uhhcui%
\author{Sebastian Trippel}%
\email{sebastian.trippel@cfel.de}%
\cfeldesy\uhhcui%
\author{Terry Mullins}\cfeldesy%
\author{Karol Długołęcki}\cfeldesy%
\author{Rosario Gonz{\'a}lez-F{\'e}rez}%
\email{rogonzal@ugr.es}%
\granada%
\author{Jochen Küpper}%
\email{jochen.kuepper@cfel.de}%
\homepage[\\\hspace*{0.8em}website:~]{https://www.controlled-molecule-imaging.org}%
\cfeldesy\uhhcui\uhhphys%
\date{\today}%
\begin{abstract}\noindent%
   We experimentally demonstrated strong adiabatic mixed-field orientation $\Nuptot=0.882$ of
   carbonyl sulfide molecules (OCS) in their absolute ground state. OCS was oriented in combined
   non-resonant laser and static electric fields inside a two-plate velocity map imaging
   spectrometer. The transition from non-adiabatic to adiabatic orientation for the rotational
   ground state was studied by varying the applied laser and static electric field. Above static
   electric field strengths of 10~kV/cm and laser intensities of $10^{11}~\SI{}{\wsqcm}$ the
   observed degree of orientation reached a plateau. These results are in good agreement with
   computational solutions of the time-dependent Schrödinger equation.
\end{abstract}
\pacs{37.10.-x, 37.10.Vz, 82.20.Bc}%
\maketitle%
\noindent%

\section{Introduction}
Molecular samples with directional order, \eg, oriented molecules, enable the extraction of
information directly in the molecular frame, for instance, from photoelectron angular
distributions~\cite{Meckel:Science320:1478, Bisgaard:Science323:1464, Holmegaard:NatPhys6:428,
   Kelkensberg:PRA84:051404, Boll:PRA88:061402}, high-order harmonic
generation~\cite{Itatani:Nature432:867, Vozzi:NatPhys7:822, Kraus:PRL113:023001}, electron and x-ray
diffractive imaging~\cite{Hensley:PRL109:133202, Kuepper:PRL112:083002, Yang:NatComm7:11232}, and
stereochemistry experiments~\cite{Brooks:Science193:11, Loesch:JCP93:4779,
   Rakitzis:Science303:1852}. For diffraction experiments, data has often to be recorded and
averaged over many shots. If the molecules in an ensemble have directional order, a molecular-frame
diffraction pattern corresponding to the single-molecule signal above noise can be
obtained~\cite{Stern:FD171:393, Filsinger:PCCP13:2076, Spence:PRL92:198102}.

Various approaches have been developed to generate oriented molecules, including brute-force
orientation using strong dc electric~\cite{Loesch:JCP93:4779, Friedrich:Nature353:412,
   Block:PRL68:1303} and magnetic~\cite{Slenczka:PRL72:1806} fields,
shaped~\cite{Ghafur:NatPhys5:289, Goban:PRL101:013001} and two-color~\cite{Vrakking:CPL271:209,
   De:PRL103:153002, Kraus:PRL113:023001} near-infrared laser pulses, terahertz
pulses~\cite{Harde:PRL66:1834, Machholm:PRL87:193001, Fleischer:PRL107:163603,
   Egodapitiya:PRL112:103002}, multi-pulse schemes~\cite{Kraus:PRL113:023001,
   Egodapitiya:PRL112:103002}, and mixed laser and dc electric field
orientation~\cite{Friedrich:JCP111:6157, Holmegaard:PRL102:023001, Ghafur:NatPhys5:289,
   Filsinger:JCP131:064309, Trippel:MP111:1738}. In mixed-field orientation, the non-resonant laser
field creates near degenerate doublets that are efficiently oriented by the dc electric
field~\cite{Friedrich:JCP111:6157, Holmegaard:PRL102:023001}. However, for the two components of the
doublet the dipole moments, and thus the molecules, point in opposite directions, resulting in a
reduced or vanishing macroscopic orientation depending on the populations of the two
components~\cite{Filsinger:JCP131:064309}. Quantum-state selection allows for the preparation of
ensembles of molecules all in a single rovibronic state~\cite{Stern:ZP39:751, Reuss:StateSelection,
   Nielsen:PCCP13:18971, Putzke:PCCP13:18962, Chang:IRPC34:557}. Strong orientation can be achieved
if these populations can be adiabatically transferred to the oriented field-dressed states.
According to the adiabaticity theorem~\cite{Born:ZP51:165}, the field-dressed dynamics are adiabatic
if a molecule remains in its eigenstate as the field strength, \eg, the laser intensity, is changed.
This condition can be fulfilled for most quantum mechanical systems, including non-resonant
adiabatic alignment~\cite{Ortigoso:PRA86:032121}, when the Hamiltonian evolves sufficiently slowly
in time. However, in contrast to adiabatic alignment, adiabatic mixed-field orientation tends to be
more challenging, because the energy-spacing within the doublets becomes very small inside the laser
field~\cite{Nielsen:PRL108:193001, Trippel:PRL114:103003}.

Here, we present a combined experimental and computational investigation of the degree of
orientation of rovibronic-ground-state-selected OCS molecules for various laser intensities and dc
electric field strengths. The presented experimental setup allows for the use of comparably strong
dc electric fields of 20~kV/cm. The experimental findings are compared to the theoretical
description obtained by solving the time-dependent Schrödinger equation for the mixed-field
orientation of the populated states within the state-selected molecular beam.

\section{Methods}
\subsection{Experiment}
\begin{figure}
   \centering
   \includegraphics[width=\linewidth]{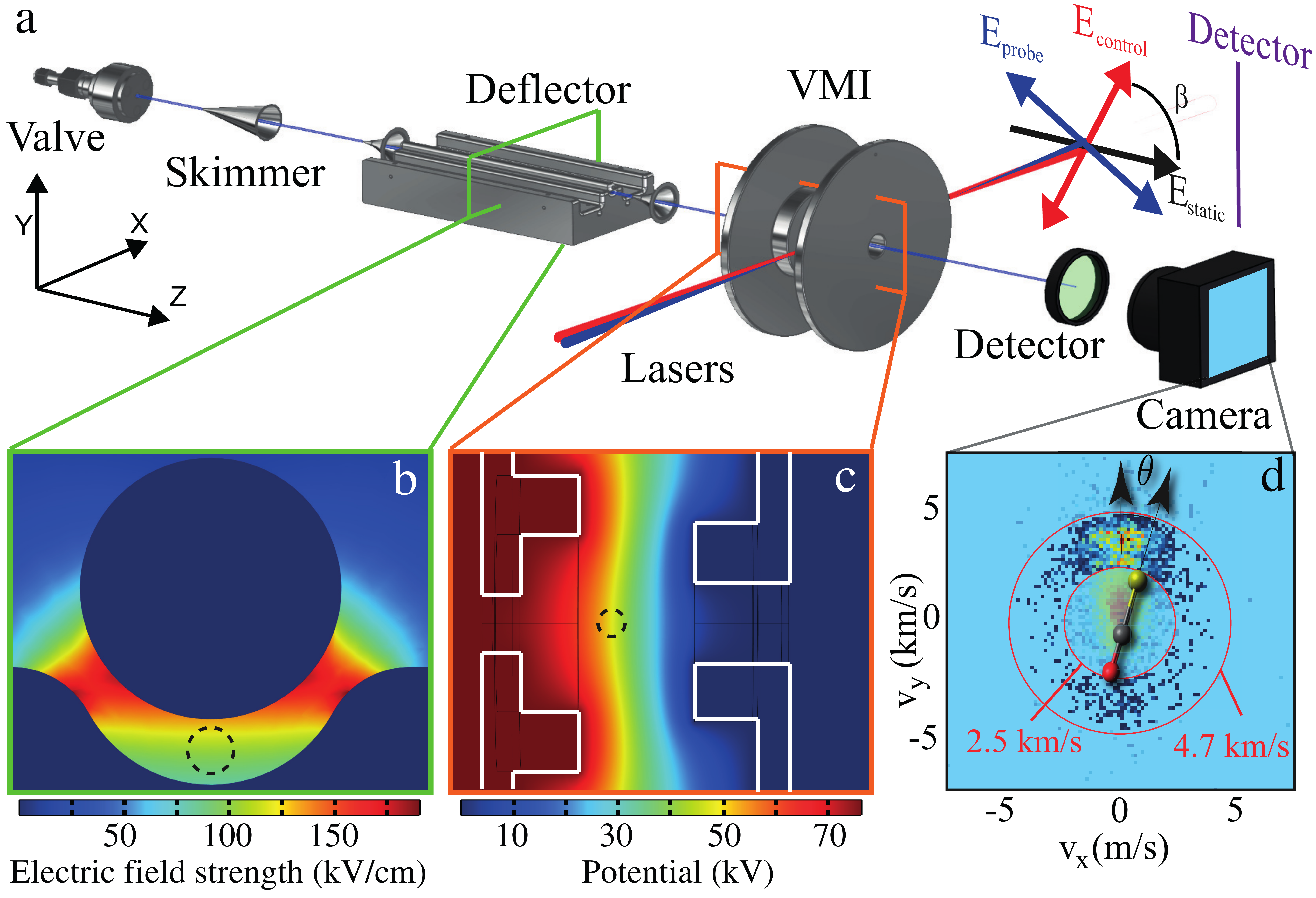}
   \caption{(Color online) (a) Experimental setup and definition of the axis system and relevant
      directions and angles. $\beta$ is the angle between the static electric field and the
      polarization vector of the control laser pulse; probe and control laser polarization are
      perpendicular to each other. Colored wireframes mark the positions of the cross sections
      depicted below. (b) Cross section of the deflector and the corresponding electric field
      strength. The position of the molecular beam is marked as a black dashed circle. (c) Cross
      section of the VMI and the corresponding electric potential. The dashed circle marks the
      interaction region of the molecular beam with the laser pulses and the white frames the
      electrodes. (d) Typical velocity map of S$^+$ ions from ionization of oriented OCS. \thetatwoD
      is defined by the angle between the laboratory fixed $Y$-axis and the molecule fixed $z$-axis
      projected onto the detector surface. Velocity cuts used in the analysis are illustrated by red
      circles in the detector image.}
   \label{fig:experimental_setup}
\end{figure}
The experimental setup is depicted in~\autoref{fig:experimental_setup}. A supersonic molecular beam
was generated by expanding a mixture of 500~ppm OCS seeded in \SI{50}{\bar} of helium into vacuum
through an Even-Lavie-valve~\cite{Even:JCP112:8068} at a repetition rate of \SI{250}{\Hz}. The beam
was collimated by two skimmers, \SI{7.4}{\centi\meter} and \SI{23.1}{\centi\meter} downstream from
the nozzle. An electrostatic deflector~\cite{Filsinger:ACIE48:6900, Chang:IRPC34:557}, placed
\SI{25.9}{\centi\meter} downstream from the nozzle, dispersed the molecular beam according to its
quantum states~\cite{Filsinger:JCP131:064309, Chang:IRPC34:557}. A cross section of the deflector
and its electric field are shown in \autoref{fig:experimental_setup}~b. A third skimmer was
positioned \SI{1.4}{\centi\meter} behind the deflector for further differential pumping.

The molecules are oriented and probed inside a velocity map imaging spectrometer (VMI) consisting of
two electrodes~\cite{Papadakis:RSI77:3101}. The control and probe laser pulses with a central
wavelength of \SI{800}{\nano\meter} were provided by an amplified femtosecond laser
system~\cite{Trippel:MP111:1738}. The temporal profile of the control laser pulse had a sawtooth
shape with a slow rising edge ($\SI{600}{\pico\second}$, 2.5-97.5\%) and a fast falling edge
($\SI{250}{\pico\second}$). We measured the spatial beam profile, in intensity, with a beam profiler
(SpiriCon SP620U), which yields $\sigma=17~\um$ and $\sigma'=18~\um$ along the two principal axes of
the profile; the first (short) axis is rotated \degree{17} away from the laboratory $Y$ axis. This
resulted in a peak intensity up to $\Icontrol\approx\SI{6E11}{\wsqcm}$. The intensity was controlled
by a half-wave plate mounted on a rotation stage in combination with a polarization filter. The
degree of orientation was probed through Coulomb explosion imaging following multiple ionization of
OCS by a 30~fs laser pulse focused down to
$\sigma_1=\SI{13}{\micro\meter}, \sigma_2=\SI{17}{\micro\meter}$, which resulted in a peak intensity
of $\Iprobe\approx\SI{1E14}{\wsqcm}$. The first principal axis is rotated by \ang{43} towards the
laboratory $Y$ axis. The relative timing between the control and probe laser pulses was varied with
a delay stage positioned in the laser beam path of the probe laser. Both laser pulses were linearly
polarized with polarization vectors perpendicular to each other. The orientation of both
polarization vectors around the laboratory fixed $X$-axis, and, therefore, the angle $\beta$ between
the ac and the dc electric fields were simultaneously controlled by a half wave plate. At
$\beta=\ang{35}$ we achieved the best compromise between the strongest orientation, with its maximum
at $\beta=\ang{0}$, and the ideal detection efficiency, with its maximum at $\beta=\ang{90}$. For
$\beta=\degree{35}$, the component of the dc electric field parallel to the polarization axis of the
control laser is reduced to $\cos{(\ang{35})\,\Estat\approx0.82\,\Estat}$.

The temporal intensity profile of the chirped control laser pulse has been determined as described
in appendix~\ref{sec:laser-pulse} and is shown in \autoref{fig:hv-pico-time}~c. For use in
computations we applied a Fourier-transform-based low-pass filter to remove frequencies above 81~GHz
in the temporal profile of the pulse, which correspond to the limit of the pixel size of the CCD
within the spectrometer; the resulting pulse is depicted by the blue area in \autoref{fig:delay}~a.

The deflection of the molecular beam was characterized by vertically scanning the $Y$ position of
the focus of the probe laser pulse across the molecular beam using a corresponding translation of
the focusing lens. The integrated ion signal at each position is proportional to the column density
at the corresponding $Y$ position in the molecular beam.
\begin{figure}
   \centering
   \includegraphics[width=\linewidth]{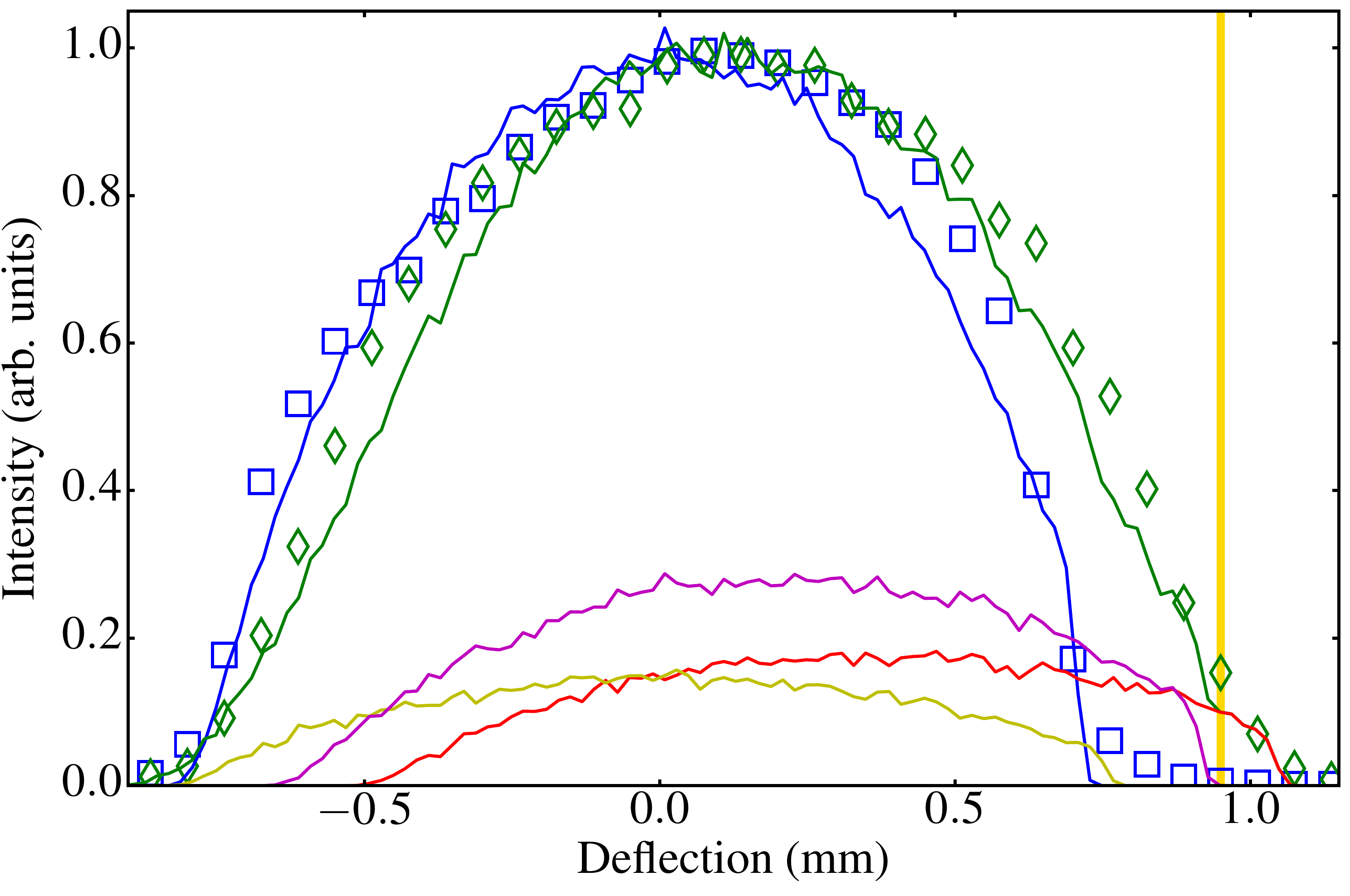}
   \caption{(Color online) Measured profiles along the laboratory $Y$-axis for the deflected
      (rhombuses) and undeflected (squares) molecular beams. The solid lines represent simulations
      of the density profile for the deflected \ket{0,0}, \ket{1,0}, \ket{1,1} states in red, brown,
      and purple, respectively and for the sum of these states in green. The blue line is the sum of
      the undeflected states. The vertical yellow line depicts the position of the laser focus used
      for the orientation experiments in the molecular beam.}
   \label{fig:deflection}
\end{figure}
\autoref{fig:deflection} shows the measured normalized density profiles of the deflected (rhombuses)
and the undeflected (squares) molecular beams. The molecules are deflected to positive $Y$ values by
the interaction of their quantum-state-specific dipole moment \mueff with the inhomogeneous electric
field. The solid lines represent numerical simulations~\cite{Chang:CPC185:339, Chang:IRPC34:557} of
the density profiles for the undeflected (blue) and deflected (green) molecular beams as well as the
individual contributions to the deflected beam by the $\ket{J,M}=\ket{0,0}$, \ket{1,0}, \ket{1,1}
states in red, brown, and purple, respectively. From these simulations we obtain a rotational
temperature of 2~K for the original molecular beam; this fairly high
temperature~\cite{Chang:IRPC34:557} is ascribed to the low stagnation pressure and not fully
optimized operation conditions of the valve, but is not critical for the investigation performed
here. The yellow bar at \SI{0.95}{\mm} in \autoref{fig:deflection} indicates the position where the
orientation experiments were performed. The estimated ground state population at this position was
$0.95\pm0.05$, with the remaining population in the \ket{1,1} state.

It was predicted that dc field strengths on the order of $\Estat=\SI{10}{\kvcm}$ are required to
achieve adiabatic orientation with 500~ps pulses~\cite{Nielsen:PRL108:193001}. In order to achieve
such strong fields a two plate velocity map imaging spectrometer~\cite{Papadakis:RSI77:3101} was set
up. A sectional view of the electrodes and typical potentials are shown in
\autoref{fig:experimental_setup}\,c. The two-plate design allows for a much stronger field for a
given repeller-electrode potential than for the classical three-plate VMI~\cite{Eppink:RSI68:3477},
\eg, $\Estat=\SI{20.7}{\kvcm}$ at $U_{r}=$\SI{80}{\kV}. In addition, as the magnification of the
velocity map scales with $1/\sqrt{\text{U}_{\text{r}}}$, the measured detector images are larger
compared to a classical VMI operated at the same electric field strength. Velocity-focusing
conditions were obtained by positioning the laser focus at a specific $Z$ position between the two
electrodes. This position is independent of the applied repeller voltage, which allows for
continuous tuning of \Estat without a change of the VMI focusing conditions.

The velocity maps are detected by a position sensitive detector, a combination of two multi-channel
plates (MCPs) in Chevron configuration and a phosphor screen. A CMOS camera (Optronis CL600X2) with
a repetition rate of \SI{1}{\kHz} was used to film the screen. The positions of individual ions were
determined via a centroiding algorithm~\cite{Wester1998}. Gating the detector by a fast high-voltage
switch (Behlke HTS 31-03-GSM) allowed to distinguish ionic fragments by their time of flight and to
record VMIs for individual fragments. \autoref{fig:experimental_setup}\,d shows a typical
S$^+$-position histogram. Red circles indicate the area between
$v_{\parallel}=\sqrt{v^2_x + v^2_y}=\SI[per-mode=symbol]{2500}{\meter\per\second}$ and
$v_{\parallel}=\SI[per-mode=symbol]{4700}{\meter\per\second}$; this range was used to determine the
degree of orientation in all measurements. Ions recorded in this area originate from the Coulomb
fragmentation channel $\text{OCS}+n\gamma\rightarrow\text{OC}^{+}+\text{S}^+$, which is a
directional fragmentation along the C-S bond of the molecule. Slower fragmentation channels with
velocities $v_{\parallel}<\SI[per-mode=symbol]{2500}{\meter\per\second}$, resulting from
singly-ionized molecules, are much more intense and not shown~\cite{Trippel:PRL114:103003}. The
degree of orientation is characterized by the ratio $\Nuptot$ of S$^+$ ions hitting the detector on
the upper half $\text{N}_{\text{up}}$ divided by the total number of S$^+$ ions
$\text{N}_{\text{tot}}$.

\subsection{Theory}
\label{ssec:methods:theoretical}
To obtain physical insight into the experimental orientation, a theoretical description of the
rotational dynamics of OCS in the experimental field configuration was performed. The dc electric
field is always turned on adiabatically, which was computationally checked to be valid for the
current experimental parameters, and the adiabatic pendular states of the dc-field configuration
were taken as the initial states in these calculations. Then, the time-dependent Schr\"odinger
equation was solved for a constant dc field using the temporal profile of the experimental control
laser pulses. To compare with the experimental observations, the theoretical orientation ratio
\Nuptot was computed including the volume effect, which took into account the spatial intensity
profiles of the control and the probe laser pulses (\emph{vide supra}) and the experimental
velocity distribution of the ions after the Coulomb explosion in the range
$2500\text{~m/s}\le{}v_{\parallel}\le4700\text{~m/s}$~\cite{Omiste:PCCP13:18815}. If the mixed-field
dynamics was adiabatic, the molecule remained in the same pendular eigenstate as the laser pulse is
turned on and the Hamiltonian evolves with time~\cite{Born:ZP51:165}. The rotational dynamics were
analyzed by projecting the time-dependent wave function on the basis formed by the adiabatic
pendular states, which was obtained by solving the time-independent Schr\"odinger equation for the
instantaneous Hamiltonian at time $t$, including both, the interactions with the ac and dc fields.
These projections onto the adiabatic pendular basis allowed us to disentangle the field-dressed
dynamics for each state of the molecular beam as well as to identify the sources of non-adiabatic
effects.

\section{Experimental Results}
\label{sec:results}
\begin{figure}
   \centering
   \includegraphics[width=\linewidth]{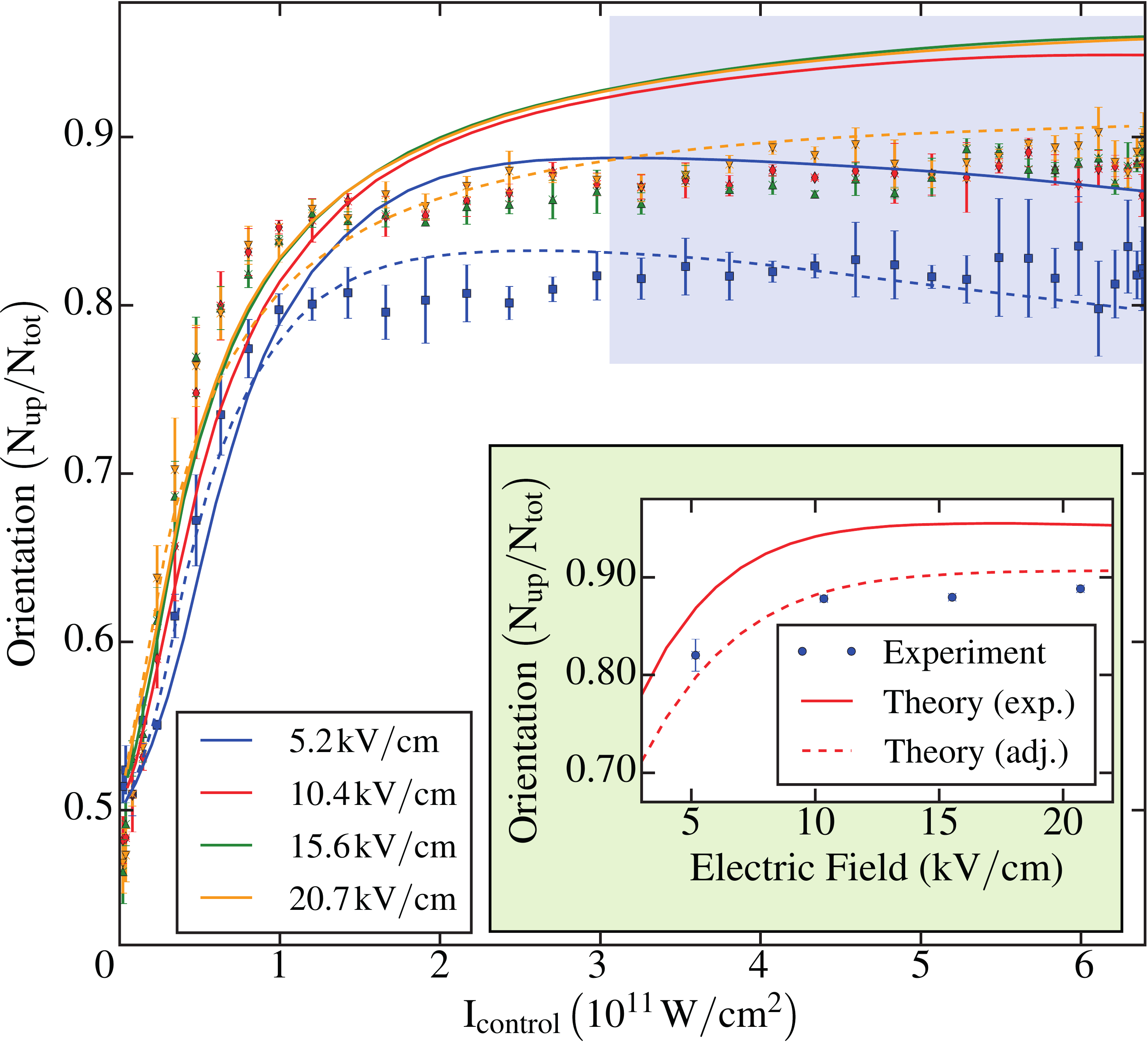}
   \caption{(Color online) Experimental (points) and theoretical (solid and dashed lines) degree of
      orientation as a function of the peak control laser intensity for $\beta=\degree{35}$ and
      static electric field strengths of 5.2~kV/cm, 10.4~kV/cm, 15.6~kV/cm, and 20.7~kV/cm. In the
      inset the experimental (points) and theoretical (solid line) mean degree of orientation,
      averaged over control laser intensities between $\Icontrol=\SI{3e11}{\wsqcm}$ and
      \SI{6.4E11}{\wsqcm} (blue area of the main graph), is shown as a function of the dc
      electric field \Estat. For the theoretical degree of orientation, depicted by the solid red
      line, we assumed the populations obtained from the simulated deflection profiles shown in
      \autoref{fig:deflection}. The dashed line represents calculations with an adjusted population
      distribution as described in the text.}
   \label{fig:intensity-35}
\end{figure}
\autoref{fig:intensity-35} shows the experimental degree of orientation $\Nuptot$ as a function of
the peak control laser intensity \Icontrol for experimental dc electric field strengths
$\Estat=5.2$~kV/cm, 10.4~kV/cm, 15.6~kV/cm, and 20.7~kV/cm. A small degree of
anti-orientation\footnote{For an anti-oriented state, it holds that
$-1\le \oricost < 0$ and $0 \le \Nuptot < 0.5$.} at zero and weak
control laser intensities was observed. This is due to the combined effect of ``brute-force'' orientation, generated by the static electric field of the VMI spectrometer, and geometric alignment, \ie, selective ionization of OCS by the probe laser, which
results in preferred ionization of anti-oriented molecules. For all four electric field strengths,
the degree of orientation increased with increasing peak control laser intensity up to
$\Icontrol\approx\SI{1E11}{\wsqcm}$. The slope of the experimental degree of orientation was the
same for $\Estat=10.4$, 15.6, and \SI{20.7}{\kvcm}, while it was slightly lower for \SI{5.2}{\kvcm}.
For $\Icontrol>\SI{2E11}{\wsqcm}$, the degree of orientation was nearly constant with further
increasing laser intensities. In this plateau region the degree of orientation was, within error
estimates, independent of the dc field strengths for $\Estat\ge10$~kV/cm, while it was reduced for
\SI{5.2}{\kvcm}. The inset of \autoref{fig:intensity-35} shows the experimental and theoretical mean
degree of orientation in the plateau region as a function of the dc field strength. All data points
between $\Icontrol=\SI{3e11}{\wsqcm}$ and \SI{6.4e11}{\wsqcm}, indicated by the blue area in the
main figure, were taken into account. The experimentally obtained orientation was
$\Nuptot=0.882\pm0.004$ for dc field strengths of \SI{10.4}{\kvcm} and above, indicating nearly
adiabatic orientation for the ground state. For \SI{5.2}{\kvcm}, we had a 7\,\% smaller degree of
orientation of $\Nuptot=0.820$.

\section{Discussion}
To understand the saturation of the degree of orientation as a function of the dc field strength the
eigenenergies of the adiabatic pendular states of OCS were examined.
\begin{figure}
   \centering
   \includegraphics[width=\linewidth]{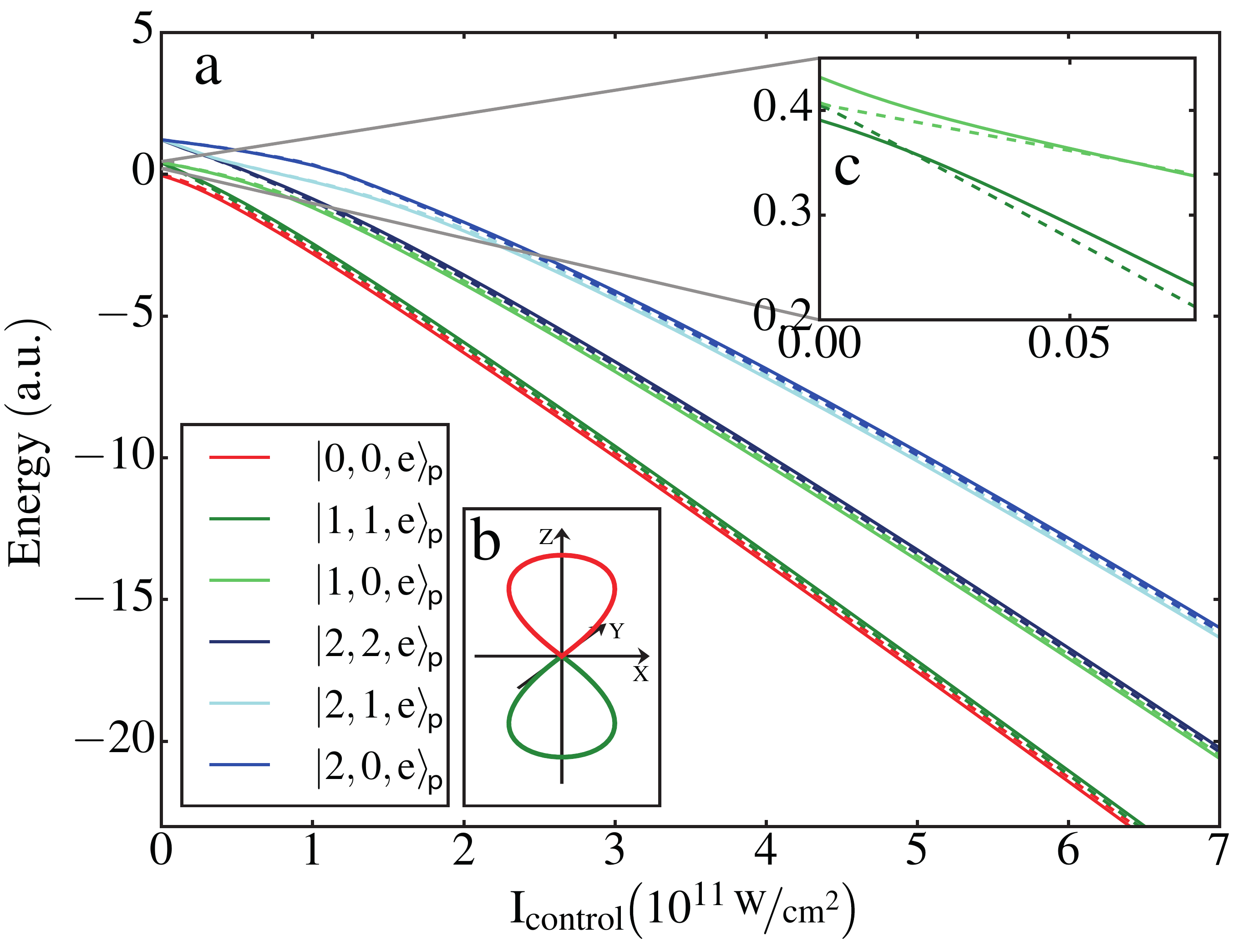}
   \caption{(Color online) (a) Energy of the \pket{0,0,e}, \pket{1,1,e}, \pket{1,0,e}, \pket{2,2,e},
      \pket{2,1,e}, and \pket{2,0,e} states as a function of the control laser intensity \Iprobe for
      a static electric field strengths with $\Estat=\SI{5.2}{\kvcm}$ (dashed lines), and
      $\Estat=\SI{20.7}{\kvcm}$ (solid lines). (b) Sketch of the square of the \pket{0,0,e} and
      \pket{1,1,e} wave functions at $\Estat=\SI{20.7}{\kvcm}$.}
   \label{fig:theory}
\end{figure}
\autoref{fig:theory}~a shows the eigenenergies for $\Estat=\SI{5.2}{\kvcm}$ (dashed lines) and
\SI{20.7}{\kvcm} (solid lines) as a function of the control laser intensity.\footnote{The adiabatic
   pendular states \pket{J,M,s} are labeled by the quantum numbers $J$ and $M=|m_J|$ of the
   adiabatically correlated field-free rotor states \ket{J,M} and the symmetry $s$ with respect to
   reflection at the plane defined by the dc and ac electric field vectors, denoted $e$ for even and
   $o$ for odd. For the adiabatic following, we take into account that the dc field is turned on
   first. Once the maximum dc electric field strength is achieved, the ac laser pulse is turned on.
   These adiabatic-pendular-state labels are amended by a subscript $p$.} As the laser intensity
increases, the two pendular states \pket{0,0,e} and \pket{1,1,e} form a near-degenerate doublet;
their energy spacing in the strong-ac-field limit is given by
$\Delta{E}\approx2\mu_\text{p}\Estat\cos{\beta}$ with the permanent dipole moment $\mu_\text{p}$. In
order to ensure adiabatic orientation, any time scale contained in the temporal envelope of the
control laser pulse has to be longer than the time scale that corresponds to the instantaneous
energy difference in the laser field. Therefore, \eg, the rise time of the control laser pulse has
to be longer for a dc field of $\Estat = \SI{5.2}{\kvcm}$ than for a dc field of \SI{20.7}{\kvcm}.
If this requirement is not fulfilled, the dynamics becomes non-adiabatic and population is
transferred between the two pendular states forming the doublet. In this case the resulting
orientation for a system starting in the ground state is reduced since the two pendular states
orient in opposite directions, see \autoref{fig:theory}~b.
For rotationally excited states, the field-dressed dynamics is more complicated. In addition to the
pendular doublet formation, the field-free-degenerate \ket{J,M} manifold splits into distinct $M$
components. This results in narrow avoided crossings between energetically neighboring pendular
states in the combined field. The avoided crossing between the \pket{1,1,e} and \pket{1,0,e} states
is shown in \autoref{fig:theory}~c for $\Estat=\SI{5.2}{\kvcm}$ (dashed lines) and \SI{20.7}{\kvcm}
(solid lines). A larger splitting is observed for stronger dc fields and, therefore, the
corresponding dynamics is more adiabatic. Our calculations have shown that a field strength on the
order of \SI{400}{\kvcm} is required to provide adiabatic orientation for the $J=1$ states with a
control pulse similar to the experimental one, but without roughness.

In \autoref{fig:intensity-35} the slopes of all experimentally determined degrees of orientation
were steeper than for the calculated ones. A possible reason for this discrepancy are errors in the
experimentally determined intensity of the control laser, which relies on a determination of the
spatial profile of the laser focus. For these beam-profile measurements, the laser beam had to be
attenuated by seven orders of magnitudes, which was achieved by using reflections of optical flats
and neutral density filters and which might have affected the beam profile. Furthermore, the profile
of the laser focus might change slightly with pulse energy, for instance, due to self focusing at
high pulse energies. The slopes of the experiment and the calculations match nicely if we assume a
1.43 times more intense control laser beam, well within our error estimates.

\begin{figure}
   \centering
   \includegraphics[width=\linewidth]{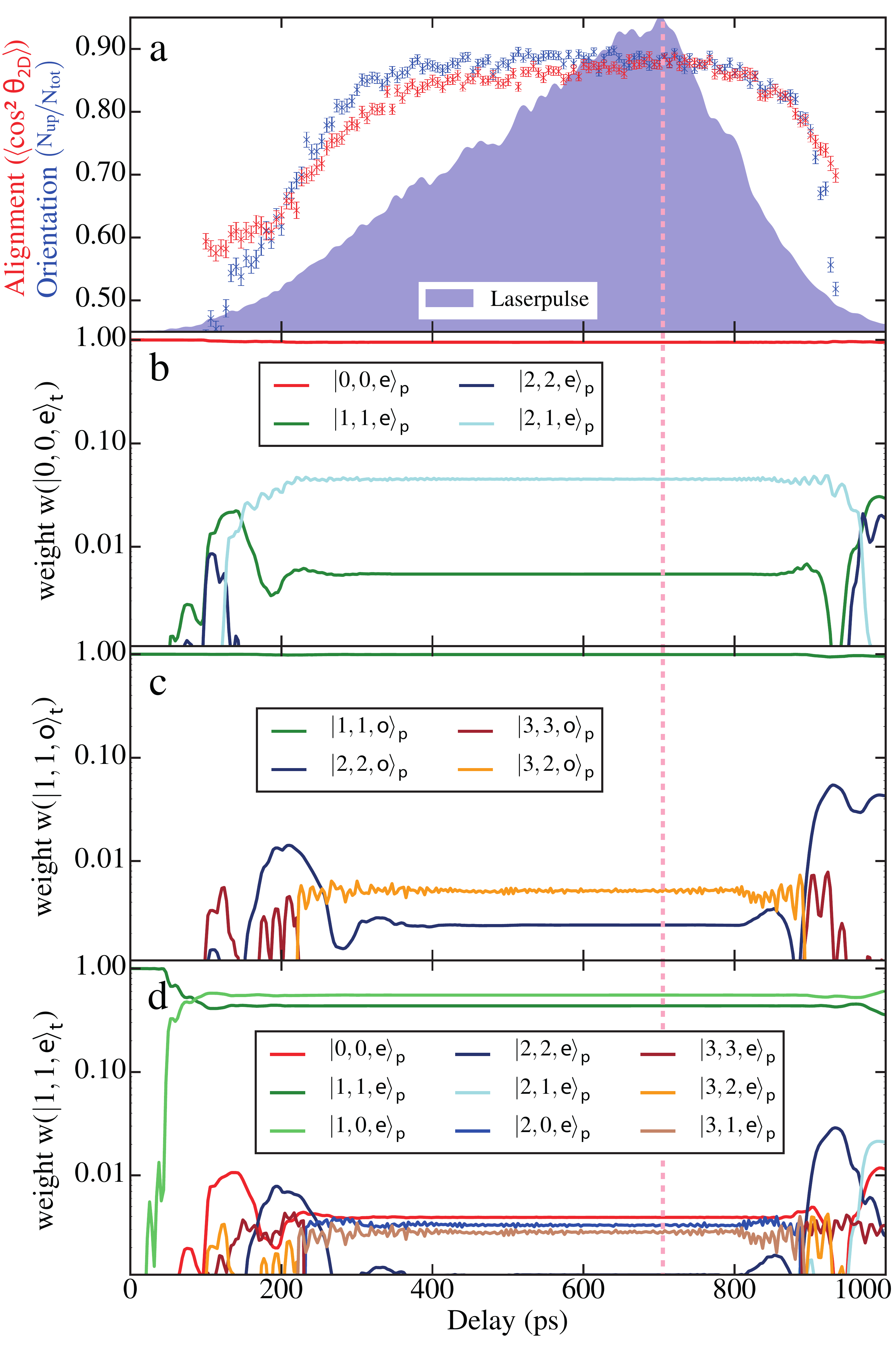}
   \caption{(Color online) (a) Degree of orientation and alignment as a function of the delay
      between the control and the probe laser with a peak intensity of the control laser
      $\Icontrol=\SI{6E11}{\wsqcm}$, a static electric field of $\Estat=\SI{20.7}{\kvcm}$, and a
      laser polarization direction of $\beta=\ang{35}$. The purple area depicts the temporal profile
      of the control laser. (b,\,c,\,d) weights of the basis functions $\pket{J,M,s}$ of the
      time-dependent wave functions \tket{0,0,e}, \tket{1,1,o} and \tket{1,1,e}, respectively. The
      dashed line marks the delay that we used for the measurements shown in
      \autoref{fig:delay}}
   \label{fig:delay}
\end{figure}

The theoretical degree of orientation taking into account our experimental conditions are shown as
solid lines in \autoref{fig:intensity-35}. The initial rotational state distribution is given by
$w(\pket{0,0,e})=0.95$, $w(\pket{1,1,e})=0.025$, and $w(\pket{1,1,o})=0.025$, corresponding to the
state distribution obtained from the deflection profile in \autoref{fig:deflection} assuming that
the individual states are adiabatically transferred from the deflector to the interaction region
inside the velocity map imaging spectrometer. The experimentally determined temporal laser intensity
profile, shown in \autoref{fig:delay}~a, was taken into account. In comparison to the experimental
results we observe a higher degree of orientation for the theoretical curves. We attribute this
discrepancy to the following effects: First, simulations have shown that molecules in excited
rotational states are not adiabatically transferred from the deflector to the velocity map imaging
spectrometer. The non adiabatic transfer is mostly caused by rotating electric field
vectors~\cite{Wall:PRA81:033414} in the fringe field regions of the deflector and the VMI. Moreover,
Majorana transitions could occur in field free regions. Second, a smoother temporal profile of the
laser (\emph{vide supra}) pulse would result in a larger (smaller) weight of the \pket{1,1,e}
(\pket{1,0,e}) state due to a more-adiabatic passage at the corresponding avoided crossing. Because
\pket{1,1,e} is anti-oriented while \pket{1,0,e} is oriented this would lead to a decreased degree
of orientation.

Adjusting the input parameters for the calculations, a better agreement between experiment and
theory was obtained. The dashed lines in~\autoref{fig:intensity-35} correspond to calculations where
the peak intensity was increased by a factor of 1.43 and an initial state population of
$w(\pket{0,0,e})=0.8$, $w(\pket{1,1,e})=0.066$, $w(\pket{1,1,o})=0.066$, and $w(\pket{1,0,e})=0.066$
was assumed. The results of these calculations show a better agreement with the experimental
measurements.

The theoretical curve for \SI{5.2}{\kvcm} shows a decrease of the degree of orientation for
increasing laser intensities above $3\times10^{11}$~W/cm$^2$. This can be attributed to increased
non-adiabatic coupling and population transfer between the \pket{0,0,e} and \pket{1,1,e} states,
which gets more important for higher intensities as the slope of the intensity-change and,
therefore, the temporal variation of the Hamiltonian gets faster. The theoretical decrease is within
the error of the experimental data and is not resolved in the experimental results.

The non adiabatic dynamics can be further investigated by studying the time dependent weights of the
decomposition of the time-dependent wave function $\tket{J,M,s}$ in the basis of the adiabatic
pendular states $\pket{J,M,s}$.
The purple area in \autoref{fig:delay}\,a depicts the temporal laser beam profile with its slow
rising and a fast falling edge obtained from the spectrum of the chirped control laser pulse. In
addition, the experimental degree of orientation and alignment as a function of the relative timing
between the control laser pulse and the probe pulse is shown. The dashed line mark the delay we used
for the intensity measurements. Both, the degree of orientation and alignment peak at this
intensity. The intensity used for the calculation of the time dependent weights is given by
$\SI{6e11}{\wsqcm}$. The dc field strengths is $\SI{20.7}{\kvcm}$. In~\autoref{fig:delay}\,b--d the
time dependent weights $w$ of systems initially in the $\tket[=0]{0,0,e}=\pket{0,0,e}$,
$\tket[=0]{1,1,o}=\pket{1,1,o}$, and $\tket[=0]{1,1,e}=\pket{1,1,e}$ states are presented. A
deviation from a weight $w=1$ indicates population transfer and, therefore, non-adiabatic dynamics.

The field-dressed dynamics of the ground state \tket{0,0,e} in \autoref{fig:delay}\,b is
characterized by the formation of the pendular pair at a delay of approximately 50~ps, when
population is transferred to the first rotational excited state resulting in a weight
$w(\pket{1,1,e})=5.5\cdot10^{-3}$. In addition, population transfer to states that correlate
adiabatically to the $J=2$ manifold is observed in the calculation. This is attributed to the
roughness of the time-profile and the sudden changes in intensity of the experimental pulse, and not
due to the slow $\SI{600}{\pico\second}$ rise time of the laser pulse, which is long compared to the
rotational period of 82~ps. Initially, population is transferred to \pket{2,2,e}. At stronger ac
fields, this state encounters an avoided crossing at which practically all population is
diabatically transferred to the \pket{2,1,e} state. Furthermore, for stronger fields further
population transfer from the ground-state pendular doublet proceeds to this \pket{2,1,e} state,
which reaches a population of $w(\pket{2,1,e})\approx4.5\cdot10^{-2}$. Although the field-dressed
dynamics of \tket{0,0,e} including the experimental laser profile is non-adiabatic, the pendular
states contributing to the dynamics, \pket{0,0,e}, \pket{1,0,e} and \pket{2,1,e}, are all strongly
oriented and a computed degree of orientation of $\Nuptot=0.976$ is obtained for \tket{0,0,e} at the
peak laser intensity. We have also performed calculations with an ideal pulse without roughness,
which was generated by fitting error functions to the experimental pulse. For this completely smooth
theoretical pulse the field-dressed dynamics of the ground-state state is only affected by the
formation of the pendular doublet and the dynamics would be completely adiabatic, confirming that
the remaining non-adiabaticity is due to the, albeit small, experimental noise in the laser
intensity.

For the ground state of the odd irreducible representation, \tket{1,1,o}, we encounter an equivalent
field-dressed dynamics, shown in \autoref{fig:delay}~c; it is slower because the pendular doublet
formation occurs at stronger ac fields. At approximately $100$~ps we observe an ac-field-induced
population transfer between the initially populated \pket{1,1,o} state and the \pket{3,3,o} state.
At $200$~ps later this population transfers diabatically to the \pket{3,2,o} state. Almost
simultaneously, and due to the formation of the first pendular doublet in this irreducible
representation, there is some population transferred from the \pket{1,1,o} state to \pket{2,2,o}. At
the peak intensity, the pendular states significantly contributing to the time-dependent wave
function \tket{1,1,o} are $w(\pket{1,1,o})=0.993$, $w(\pket{2,2,o})=2.4\cdot10^{-3}$, and
$w(\pket{3,2,o})=5.2\cdot10^{-3}$. Analogously to the absolute ground-state, the dynamics of this
state \tket{1,1,o} in an ideal pulse would only be affected by the formation of the pendular doublet
between the states \pket{1,1,o} and \pket{2,2,o}.

In~\autoref{fig:delay}\,d the weights of the expansion coefficients of the time-dependent
\tket{1,1,e} wave function for the experimental pulse with peak intensity $\SI{6e11}{\wsqcm}$ and a
dc field with strength $\SI{20.7}{\kvcm}$ is presented. The mixed-field dynamics of this state is
more complicated. In the presence of a tilted static electric field, the pendular states
\pket{1,1,e} and \pket{1,0,e} are energetically very close, and even a weak ac field provokes a
strong coupling between them. At weak laser intensities, when the splitting of the \pket{1,M,e}
manifold due to the strong dc field takes place, a significant amount of population is transferred
between them. Thereafter, the state \pket{1,0,e} possesses the dominant contribution to the
\tket{1,1,e} state. This non-adiabatic behavior takes place before the pendular doublet formation,
\ie, at low intensities and short times. By further increasing the ac field strength, population is
transferred to \pket{0,0,e} and \pket{2,2,e} due to the formation of the pendular doublets with
\pket{1,1,e} and \pket{1,0,e}, respectively. Due to ac-field-induced couplings with other pendular
states and narrow avoided crossings, we observed that many adiabatic pendular states contribute to
the dynamics. At the peak intensity, the pendular states that significantly contribute to the
time-dependent wave function \tket{1,1,e} are $w(\pket{0,0,e})=3.9\cdot10^{-3}$,
$w(\pket{1,1,e})=0.435$, $w(\pket{1,0,e})=0.553$, $w(\pket{2,2,e})=1.1\cdot10^{-3}$,
$w(\pket{2,1})=6\cdot10^{-4}$, $w(\pket{2,0,e})=3\cdot10^{-3}$, $w(\pket{2,0,e})=3\cdot10^{-3}$, and
$w(\pket{3,1,e})=2.8\cdot10^{-3}$. Since \pket{1,1,e} is anti-oriented and cancels the contributions
form other states, \tket{1,1,e} shows no orientation at the peak intensity, \ie, $\Nuptot=0.50$. In
contrast, the dynamics of this \tket{1,1,e} state in an ideal pulse without roughness is mainly
affected by the splitting of the \pket{1,M,e} manifold, and weakly by the subsequent formation of
the pendular pairs. For this ideal pulse, only two adiabatic pendular states contribute
significantly, with $w(\pket{1,1,e})=0.652$ and $w(\pket{1,0,e})=0.346$, resulting in
anti-orientation. For the ideal pulse with peak intensity $\SI{6e11}{\wsqcm}$ and a dc field of
\SI{50}{\kvcm}, the non-adiabaticity during the splitting of the \pket{J,M,e} manifold would be
significantly reduced, and the population-transfer from \pket{1,1,e} to \pket{1,0,e} would be
$<0.011$.

\section{Conclusion}
%
Adiabatic mixed field orientation of ground-state-selected OCS molecules has been demonstrated using
strong dc electric fields of $10$--$20$~kV/cm. The experiments demonstrate strong orientation with
$\Nuptot=0.882$ in agreement with our theoretical description. For dc electric fields of
\SI{10.4}{\kvcm} or stronger, the observed degree of orientation was independent of the dc electric
field strength, which indicated that the molecules in their ground state are oriented adiabatically.
Comparison with calculations showed that only a very small fraction of the population was
transferred to excited states. The deviation of the degree of orientation from its maximal possible
value of $\Nuptot=0.976$ was attributed to contributions of excited rotational states present in the
deflected part of the molecular beam that was used in the orientation experiments. Preparing the
molecules in the absolute ground state would result in full adiabatic orientation dynamics and,
therefore, in an even higher degree of orientation. The adiabatic orientation of an excited
rotational state is more challenging. Avoided crossing and the degeneracy at low laser intensity
make non-adiabatic behavior more likely.

Compared to other techniques~\cite{Ghafur:NatPhys5:289, Goban:PRL101:013001, Vrakking:CPL271:209,
   De:PRL103:153002, Kraus:PRL113:023001} our approach ensures a strong degree of orientation while
employing only moderately strong laser intensities on the order of \SI{1e11}{\wsqcm}, which are far
below the onset of ionization, even for larger molecules~\cite{Strohaber:PRA84:063414}. Moreover,
these moderate fields strengths allow for the investigation of chemical dynamics, using
molecular-frame-imaging approaches, without significant distortions of the dynamics.

Our findings hold for polar molecules in general, as the Hamiltonian can be rescaled accordingly.
Due to the complexity of the rotational level structure of asymmetric tops, non-adiabatic effects
will have a larger impact on the orientation dynamics of these more complex
molecules~\cite{Omiste:PRA88:033416, Omiste:JCP135:064310}. Nevertheless, our finding hold for any
molecule prepared in the rotational ground state; the experimental realization of such a sample was
experimentally demonstrated for C$_7$H$_5$N using the alternating-gradient $m/\mu$
selector~\cite{Filsinger:PRL100:133003, Filsinger:PRA82:052513, Putzke:PCCP13:18962}.

The improved adiabaticity in mixed field orientation and the resulting increase in the degree of
orientation will improve, especially for complex molecules, imaging experiments with fixed-in-space
molecules such as the investigation of molecular-frame photoelectron angular
distributions~\cite{Holmegaard:NatPhys6:428, Pullen:NatComm6:7262, Zeidler:PRL95:203003} or the
recording of molecular movies by x-ray~\cite{Kuepper:PRL112:083002} and electron
diffraction~\cite{Hensley:PRL109:133202, Yang:NatComm7:11232} or photoelectron
holography~\cite{Boll:PRA88:061402, Boll:FD17171:57}.

\section{Acknowledgements}
Besides DESY, this work has been supported by the \emph{Deutsche Forschungsgemeinschaft} (DFG)
through the excellence cluster ``The Hamburg Center for Ultrafast Imaging -- Structure, Dynamics and
Control of Matter at the Atomic Scale'' (CUI, EXC1074) and the Helmholtz Association ``Initiative
and Networking Fund''. R.G.F.\ gratefully acknowledges financial support by the Spanish project
FIS2014-54497-P (MINECO) and the Andalusian research group FQM-207.

\appendix
\section{Measurement of the temporal control laser profile}
\label{sec:laser-pulse}
The temporal profiles of the control-laser pulses were deduced from a measurement of their spectrum
and a calibrated wavelength-to-time conversion. All spectra were recorded with a commercial
spectrometer (Photon Control SPM-002-X).
\begin{figure}
   \centering
   \includegraphics[width=\linewidth]{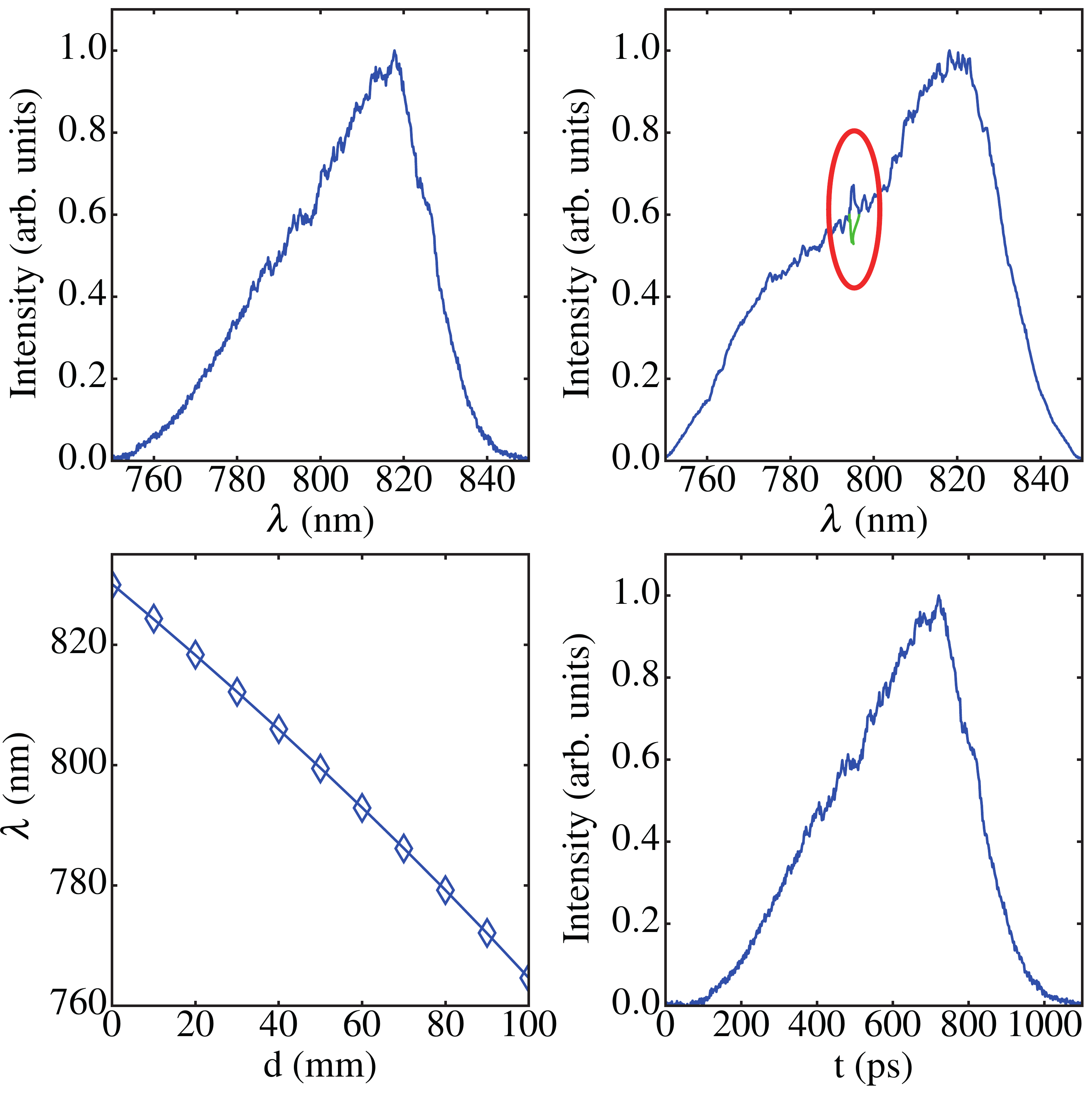}
   \caption{(Color online) (a) Spectrum of the control laser beam. (b) Interference spectrum of the
     control and probe laser beam. The red ellipse marks the position of the spectral interference
     pattern. The green line indicates the amplitude of the interference. (c) Measurement
     (rhombuses) and quadratic fit (line) of the wavelength of the interference pattern as a
     function of the delay stage position. (d) Temporal evolution of the control laser beam.}
   \label{fig:hv-pico-time}
\end{figure}
The conversion function between the measured spectrum, shown in \autoref{fig:delay}~a, and the
corresponding temporal intensity profile of the control laser pulse, shown in \autoref{fig:delay}~d,
was determined from the spectral interference of the control laser and the time-delayed $30$~fs
probe laser pulses. Both laser beams were linearly polarized, parallel to each other, colinearly
overlapped, and their simultaneous spectrum recorded. The resulting spectral interference between
the two laser pulses manifests itself as a localized large amplitude fluctuation in the spectrum.
These fluctuations mark the position ``in time'' of the short probe pulse within the spectrum of the
temporarily stretched -- chirped -- control laser pulse. As an example, the combined spectrum of
both lasers is shown for one relative timing in \autoref{fig:delay}~b. The position of the spectral
interference is highlighted by the red ellipse and the amplitude of its fluctuation is indicated by
the green curve. The wavelength $\lambda$ of the spectral interference pattern in the spectrum as a
function of the delay-stage position $d$ is shown in \autoref{fig:delay}~c. The blue line is a
quadratic fit according to $\lambda(d)=\lambda_0 + a\,d + b\,d^2$ from which we obtained
$\lambda_0=\SI[per-mode=symbol]{829.99}{\nano\meter}$,
$a=\SI[per-mode=symbol]{-0.57}{\nano\meter\per\milli\meter}$ and
$b=\SI[per-mode=symbol]{-8.65e-4}{\nano\meter\per\milli\meter\squared}$. The conversion from delay
stage position $d$ to time $t$ is given by $t=2d/c$, where $c$ is the speed of light. The factor of
two is taking into account that the light was traveling back and forth in the translation stage. The
resulting temporal profile of the control laser pulse, calculated according to
\begin{equation}
   t(\lambda) = \SI{860}{\pico\second} + \frac{2}{c} \left( \SI{698.2}{\milli\meter} -
      \SI[per-mode=fraction]{3.41}{\milli\meter\per\nano\meter} \lambda +
      \SI[per-mode=fraction]{3.098}{\milli\meter\per\nano\meter\squared} \lambda^2 \right)
   \notag
\end{equation}
is shown in \autoref{fig:hv-pico-time}~d.

\bibliography{string,cmi}

\begin{thebibliography}{59}%
\makeatletter
\providecommand \@ifxundefined [1]{%
 \@ifx{#1\undefined}
}%
\providecommand \@ifnum [1]{%
 \ifnum #1\expandafter \@firstoftwo
 \else \expandafter \@secondoftwo
 \fi
}%
\providecommand \@ifx [1]{%
 \ifx #1\expandafter \@firstoftwo
 \else \expandafter \@secondoftwo
 \fi
}%
\providecommand \natexlab [1]{#1}%
\providecommand \enquote  [1]{``#1''}%
\providecommand \bibnamefont  [1]{#1}%
\providecommand \bibfnamefont [1]{#1}%
\providecommand \citenamefont [1]{#1}%
\providecommand \href@noop [0]{\@secondoftwo}%
\providecommand \href [0]{\begingroup \@sanitize@url \@href}%
\providecommand \@href[1]{\@@startlink{#1}\@@href}%
\providecommand \@@href[1]{\endgroup#1\@@endlink}%
\providecommand \@sanitize@url [0]{\catcode `\\12\catcode `\$12\catcode
  `\&12\catcode `\#12\catcode `\^12\catcode `\_12\catcode `\%12\relax}%
\providecommand \@@startlink[1]{}%
\providecommand \@@endlink[0]{}%
\providecommand \url  [0]{\begingroup\@sanitize@url \@url }%
\providecommand \@url [1]{\endgroup\@href {#1}{\urlprefix }}%
\providecommand \urlprefix  [0]{URL }%
\providecommand \Eprint [0]{\href }%
\providecommand \doibase [0]{http://dx.doi.org/}%
\providecommand \selectlanguage [0]{\@gobble}%
\providecommand \bibinfo  [0]{\@secondoftwo}%
\providecommand \bibfield  [0]{\@secondoftwo}%
\providecommand \translation [1]{[#1]}%
\providecommand \BibitemOpen [0]{}%
\providecommand \bibitemStop [0]{}%
\providecommand \bibitemNoStop [0]{.\EOS\space}%
\providecommand \EOS [0]{\spacefactor3000\relax}%
\providecommand \BibitemShut  [1]{\csname bibitem#1\endcsname}%
\let\auto@bib@innerbib\@empty
\bibitem [{\citenamefont {Meckel}\ \emph {et~al.}(2008)\citenamefont {Meckel},
  \citenamefont {Comtois}, \citenamefont {Zeidler}, \citenamefont {Staudte},
  \citenamefont {Pavi{\v c}i{\'c}}, \citenamefont {Bandulet}, \citenamefont
  {P{\'e}pin}, \citenamefont {Kieffer}, \citenamefont {D{\"o}rner},
  \citenamefont {Villeneuve},\ and\ \citenamefont
  {Corkum}}]{Meckel:Science320:1478}%
  \BibitemOpen
  \bibfield  {author} {\bibinfo {author} {\bibfnamefont {M.}~\bibnamefont
  {Meckel}}, \bibinfo {author} {\bibfnamefont {D.}~\bibnamefont {Comtois}},
  \bibinfo {author} {\bibfnamefont {D.}~\bibnamefont {Zeidler}}, \bibinfo
  {author} {\bibfnamefont {A.}~\bibnamefont {Staudte}}, \bibinfo {author}
  {\bibfnamefont {D.}~\bibnamefont {Pavi{\v c}i{\'c}}}, \bibinfo {author}
  {\bibfnamefont {H.~C.}\ \bibnamefont {Bandulet}}, \bibinfo {author}
  {\bibfnamefont {H.}~\bibnamefont {P{\'e}pin}}, \bibinfo {author}
  {\bibfnamefont {J.~C.}\ \bibnamefont {Kieffer}}, \bibinfo {author}
  {\bibfnamefont {R.}~\bibnamefont {D{\"o}rner}}, \bibinfo {author}
  {\bibfnamefont {D.~M.}\ \bibnamefont {Villeneuve}}, \ and\ \bibinfo {author}
  {\bibfnamefont {P.~B.}\ \bibnamefont {Corkum}},\ }\bibfield  {title}
  {\enquote {\bibinfo {title} {Laser-induced electron tunneling and
  diffraction},}\ }\href {\doibase 10.1126/science.1157980} {\bibfield
  {journal} {\bibinfo  {journal} {Science}\ }\textbf {\bibinfo {volume}
  {320}},\ \bibinfo {pages} {1478--1482} (\bibinfo {year} {2008})}\BibitemShut
  {NoStop}%
\bibitem [{\citenamefont {Bisgaard}\ \emph {et~al.}(2009)\citenamefont
  {Bisgaard}, \citenamefont {Clarkin}, \citenamefont {Wu}, \citenamefont {Lee},
  \citenamefont {Ge{\ss}ner}, \citenamefont {Hayden},\ and\ \citenamefont
  {Stolow}}]{Bisgaard:Science323:1464}%
  \BibitemOpen
  \bibfield  {author} {\bibinfo {author} {\bibfnamefont {C.~Z.}\ \bibnamefont
  {Bisgaard}}, \bibinfo {author} {\bibfnamefont {O.~J.}\ \bibnamefont
  {Clarkin}}, \bibinfo {author} {\bibfnamefont {G.}~\bibnamefont {Wu}},
  \bibinfo {author} {\bibfnamefont {A.~M.~D.}\ \bibnamefont {Lee}}, \bibinfo
  {author} {\bibfnamefont {O.}~\bibnamefont {Ge{\ss}ner}}, \bibinfo {author}
  {\bibfnamefont {C.~C.}\ \bibnamefont {Hayden}}, \ and\ \bibinfo {author}
  {\bibfnamefont {A.}~\bibnamefont {Stolow}},\ }\bibfield  {title} {\enquote
  {\bibinfo {title} {Time-resolved molecular frame dynamics of fixed-in-space
  {CS}$_2$ molecules},}\ }\href {\doibase 10.1126/science.1169183} {\bibfield
  {journal} {\bibinfo  {journal} {Science}\ }\textbf {\bibinfo {volume}
  {323}},\ \bibinfo {pages} {1464--1468} (\bibinfo {year} {2009})}\BibitemShut
  {NoStop}%
\bibitem [{\citenamefont {Holmegaard}\ \emph {et~al.}(2010)\citenamefont
  {Holmegaard}, \citenamefont {Hansen}, \citenamefont {Kalh{\o}j},
  \citenamefont {Kragh}, \citenamefont {Stapelfeldt}, \citenamefont
  {Filsinger}, \citenamefont {K{\"u}pper}, \citenamefont {Meijer},
  \citenamefont {Dimitrovski}, \citenamefont {Abu-samha}, \citenamefont
  {Martiny},\ and\ \citenamefont {Madsen}}]{Holmegaard:NatPhys6:428}%
  \BibitemOpen
  \bibfield  {author} {\bibinfo {author} {\bibfnamefont {L.}~\bibnamefont
  {Holmegaard}}, \bibinfo {author} {\bibfnamefont {J.~L.}\ \bibnamefont
  {Hansen}}, \bibinfo {author} {\bibfnamefont {L.}~\bibnamefont {Kalh{\o}j}},
  \bibinfo {author} {\bibfnamefont {S.~L.}\ \bibnamefont {Kragh}}, \bibinfo
  {author} {\bibfnamefont {H.}~\bibnamefont {Stapelfeldt}}, \bibinfo {author}
  {\bibfnamefont {F.}~\bibnamefont {Filsinger}}, \bibinfo {author}
  {\bibfnamefont {J.}~\bibnamefont {K{\"u}pper}}, \bibinfo {author}
  {\bibfnamefont {G.}~\bibnamefont {Meijer}}, \bibinfo {author} {\bibfnamefont
  {D.}~\bibnamefont {Dimitrovski}}, \bibinfo {author} {\bibfnamefont
  {M.}~\bibnamefont {Abu-samha}}, \bibinfo {author} {\bibfnamefont {C.~P.~J.}\
  \bibnamefont {Martiny}}, \ and\ \bibinfo {author} {\bibfnamefont {L.~B.}\
  \bibnamefont {Madsen}},\ }\bibfield  {title} {\enquote {\bibinfo {title}
  {Photoelectron angular distributions from strong-field ionization of oriented
  molecules},}\ }\href {\doibase 10.1038/NPHYS1666} {\bibfield  {journal}
  {\bibinfo  {journal} {Nat. Phys.}\ }\textbf {\bibinfo {volume} {6}},\
  \bibinfo {pages} {428} (\bibinfo {year} {2010})},\ \Eprint
  {http://arxiv.org/abs/1003.4634} {arXiv:1003.4634 [physics]} \BibitemShut
  {NoStop}%
\bibitem [{\citenamefont {Kelkensberg}\ \emph {et~al.}(2011)\citenamefont
  {Kelkensberg}, \citenamefont {Rouz{\'e}e}, \citenamefont {Siu}, \citenamefont
  {Gademann}, \citenamefont {Johnsson}, \citenamefont {Lucchini}, \citenamefont
  {Lucchese},\ and\ \citenamefont {Vrakking}}]{Kelkensberg:PRA84:051404}%
  \BibitemOpen
  \bibfield  {author} {\bibinfo {author} {\bibfnamefont {F.}~\bibnamefont
  {Kelkensberg}}, \bibinfo {author} {\bibfnamefont {A.}~\bibnamefont
  {Rouz{\'e}e}}, \bibinfo {author} {\bibfnamefont {W.}~\bibnamefont {Siu}},
  \bibinfo {author} {\bibfnamefont {G.}~\bibnamefont {Gademann}}, \bibinfo
  {author} {\bibfnamefont {P.}~\bibnamefont {Johnsson}}, \bibinfo {author}
  {\bibfnamefont {M.}~\bibnamefont {Lucchini}}, \bibinfo {author}
  {\bibfnamefont {R.~R.}\ \bibnamefont {Lucchese}}, \ and\ \bibinfo {author}
  {\bibfnamefont {M.~J.~J.}\ \bibnamefont {Vrakking}},\ }\bibfield  {title}
  {\enquote {\bibinfo {title} {Xuv ionization of aligned molecules},}\ }\href
  {\doibase 10.1103/PhysRevA.84.051404} {\bibfield  {journal} {\bibinfo
  {journal} {Phys.\ Rev.\ A}\ }\textbf {\bibinfo {volume} {84}},\ \bibinfo
  {pages} {051404} (\bibinfo {year} {2011})}\BibitemShut {NoStop}%
\bibitem [{\citenamefont {Boll}\ \emph {et~al.}(2013)\citenamefont {Boll},
  \citenamefont {Anielski}, \citenamefont {Bostedt}, \citenamefont {Bozek},
  \citenamefont {Christensen}, \citenamefont {Coffee}, \citenamefont {De},
  \citenamefont {Decleva}, \citenamefont {Epp}, \citenamefont {Erk},
  \citenamefont {Foucar}, \citenamefont {Krasniqi}, \citenamefont {K{\"u}pper},
  \citenamefont {Rouz\'{e}e}, \citenamefont {Rudek}, \citenamefont {Rudenko},
  \citenamefont {Schorb}, \citenamefont {Stapelfeldt}, \citenamefont {Stener},
  \citenamefont {Stern}, \citenamefont {Techert}, \citenamefont {Trippel},
  \citenamefont {Vrakking}, \citenamefont {Ullrich},\ and\ \citenamefont
  {Rolles}}]{Boll:PRA88:061402}%
  \BibitemOpen
  \bibfield  {author} {\bibinfo {author} {\bibfnamefont {R.}~\bibnamefont
  {Boll}}, \bibinfo {author} {\bibfnamefont {D.}~\bibnamefont {Anielski}},
  \bibinfo {author} {\bibfnamefont {C.}~\bibnamefont {Bostedt}}, \bibinfo
  {author} {\bibfnamefont {J.~D.}\ \bibnamefont {Bozek}}, \bibinfo {author}
  {\bibfnamefont {L.}~\bibnamefont {Christensen}}, \bibinfo {author}
  {\bibfnamefont {R.}~\bibnamefont {Coffee}}, \bibinfo {author} {\bibfnamefont
  {S.}~\bibnamefont {De}}, \bibinfo {author} {\bibfnamefont {P.}~\bibnamefont
  {Decleva}}, \bibinfo {author} {\bibfnamefont {S.~W.}\ \bibnamefont {Epp}},
  \bibinfo {author} {\bibfnamefont {B.}~\bibnamefont {Erk}}, \bibinfo {author}
  {\bibfnamefont {L.}~\bibnamefont {Foucar}}, \bibinfo {author} {\bibfnamefont
  {F.}~\bibnamefont {Krasniqi}}, \bibinfo {author} {\bibfnamefont
  {J.}~\bibnamefont {K{\"u}pper}}, \bibinfo {author} {\bibfnamefont
  {A.}~\bibnamefont {Rouz\'{e}e}}, \bibinfo {author} {\bibfnamefont
  {B.}~\bibnamefont {Rudek}}, \bibinfo {author} {\bibfnamefont
  {A.}~\bibnamefont {Rudenko}}, \bibinfo {author} {\bibfnamefont
  {S.}~\bibnamefont {Schorb}}, \bibinfo {author} {\bibfnamefont
  {H.}~\bibnamefont {Stapelfeldt}}, \bibinfo {author} {\bibfnamefont
  {M.}~\bibnamefont {Stener}}, \bibinfo {author} {\bibfnamefont
  {S.}~\bibnamefont {Stern}}, \bibinfo {author} {\bibfnamefont
  {S.}~\bibnamefont {Techert}}, \bibinfo {author} {\bibfnamefont
  {S.}~\bibnamefont {Trippel}}, \bibinfo {author} {\bibfnamefont {M.~J.~J.}\
  \bibnamefont {Vrakking}}, \bibinfo {author} {\bibfnamefont {J.}~\bibnamefont
  {Ullrich}}, \ and\ \bibinfo {author} {\bibfnamefont {D.}~\bibnamefont
  {Rolles}},\ }\bibfield  {title} {\enquote {\bibinfo {title} {Femtosecond
  photoelectron diffraction on laser-aligned molecules: Towards time-resolved
  imaging of molecular structure},}\ }\href {\doibase
  10.1103/PhysRevA.88.061402} {\bibfield  {journal} {\bibinfo  {journal}
  {Phys.\ Rev.\ A}\ }\textbf {\bibinfo {volume} {88}},\ \bibinfo {pages}
  {061402(R)} (\bibinfo {year} {2013})}\BibitemShut {NoStop}%
\bibitem [{\citenamefont {Itatani}\ \emph {et~al.}(2004)\citenamefont
  {Itatani}, \citenamefont {Levesque}, \citenamefont {Zeidler}, \citenamefont
  {Niikura}, \citenamefont {P\'{e}pin}, \citenamefont {Kieffer}, \citenamefont
  {Corkum},\ and\ \citenamefont {Villeneuve}}]{Itatani:Nature432:867}%
  \BibitemOpen
  \bibfield  {author} {\bibinfo {author} {\bibfnamefont {J.}~\bibnamefont
  {Itatani}}, \bibinfo {author} {\bibfnamefont {J.}~\bibnamefont {Levesque}},
  \bibinfo {author} {\bibfnamefont {D.}~\bibnamefont {Zeidler}}, \bibinfo
  {author} {\bibfnamefont {H.}~\bibnamefont {Niikura}}, \bibinfo {author}
  {\bibfnamefont {H.}~\bibnamefont {P\'{e}pin}}, \bibinfo {author}
  {\bibfnamefont {J.~C.}\ \bibnamefont {Kieffer}}, \bibinfo {author}
  {\bibfnamefont {P.~B.}\ \bibnamefont {Corkum}}, \ and\ \bibinfo {author}
  {\bibfnamefont {D.~M.}\ \bibnamefont {Villeneuve}},\ }\bibfield  {title}
  {\enquote {\bibinfo {title} {Tomographic imaging of molecular orbitals},}\
  }\href {\doibase 10.1038/nature03183} {\bibfield  {journal} {\bibinfo
  {journal} {Nature}\ }\textbf {\bibinfo {volume} {432}},\ \bibinfo {pages}
  {867--871} (\bibinfo {year} {2004})}\BibitemShut {NoStop}%
\bibitem [{\citenamefont {Vozzi}\ \emph {et~al.}(2011)\citenamefont {Vozzi},
  \citenamefont {Negro}, \citenamefont {Calegari}, \citenamefont {Sansone},
  \citenamefont {Nisoli}, \citenamefont {De~Silvestri},\ and\ \citenamefont
  {Stagira}}]{Vozzi:NatPhys7:822}%
  \BibitemOpen
  \bibfield  {author} {\bibinfo {author} {\bibfnamefont {C.}~\bibnamefont
  {Vozzi}}, \bibinfo {author} {\bibfnamefont {M.}~\bibnamefont {Negro}},
  \bibinfo {author} {\bibfnamefont {F.}~\bibnamefont {Calegari}}, \bibinfo
  {author} {\bibfnamefont {G.}~\bibnamefont {Sansone}}, \bibinfo {author}
  {\bibfnamefont {M.}~\bibnamefont {Nisoli}}, \bibinfo {author} {\bibfnamefont
  {S.}~\bibnamefont {De~Silvestri}}, \ and\ \bibinfo {author} {\bibfnamefont
  {S.}~\bibnamefont {Stagira}},\ }\bibfield  {title} {\enquote {\bibinfo
  {title} {Generalized molecular orbital tomography},}\ }\href {\doibase
  10.1038/nphys2029} {\bibfield  {journal} {\bibinfo  {journal} {Nat. Phys.}\
  }\textbf {\bibinfo {volume} {7}},\ \bibinfo {pages} {822--826} (\bibinfo
  {year} {2011})}\BibitemShut {NoStop}%
\bibitem [{\citenamefont {Kraus}, \citenamefont {Baykusheva},\ and\
  \citenamefont {W{\"o}rner}(2014)}]{Kraus:PRL113:023001}%
  \BibitemOpen
  \bibfield  {author} {\bibinfo {author} {\bibfnamefont {P.~M.}\ \bibnamefont
  {Kraus}}, \bibinfo {author} {\bibfnamefont {D.}~\bibnamefont {Baykusheva}}, \
  and\ \bibinfo {author} {\bibfnamefont {H.~J.}\ \bibnamefont {W{\"o}rner}},\
  }\bibfield  {title} {\enquote {\bibinfo {title} {Two-pulse field-free
  orientation reveals anisotropy of molecular shape resonance},}\ }\href
  {\doibase 10.1103/PhysRevLett.113.023001} {\bibfield  {journal} {\bibinfo
  {journal} {Phys.\ Rev.\ Lett.}\ }\textbf {\bibinfo {volume} {113}},\ \bibinfo
  {pages} {023001} (\bibinfo {year} {2014})},\ \Eprint
  {http://arxiv.org/abs/1311.3923} {arXiv:1311.3923 [physics.chem-ph]}
  \BibitemShut {NoStop}%
\bibitem [{\citenamefont {Hensley}, \citenamefont {Yang},\ and\ \citenamefont
  {Centurion}(2012)}]{Hensley:PRL109:133202}%
  \BibitemOpen
  \bibfield  {author} {\bibinfo {author} {\bibfnamefont {C.~J.}\ \bibnamefont
  {Hensley}}, \bibinfo {author} {\bibfnamefont {J.}~\bibnamefont {Yang}}, \
  and\ \bibinfo {author} {\bibfnamefont {M.}~\bibnamefont {Centurion}},\
  }\bibfield  {title} {\enquote {\bibinfo {title} {Imaging of isolated
  molecules with ultrafast electron pulses},}\ }\href {\doibase
  10.1103/PhysRevLett.109.133202} {\bibfield  {journal} {\bibinfo  {journal}
  {Phys.\ Rev.\ Lett.}\ }\textbf {\bibinfo {volume} {109}},\ \bibinfo {pages}
  {133202} (\bibinfo {year} {2012})}\BibitemShut {NoStop}%
\bibitem [{\citenamefont {K{\"u}pper}\ \emph {et~al.}(2014)\citenamefont
  {K{\"u}pper}, \citenamefont {Stern}, \citenamefont {Holmegaard},
  \citenamefont {Filsinger}, \citenamefont {Rouz\'{e}e}, \citenamefont
  {Rudenko}, \citenamefont {Johnsson}, \citenamefont {Martin}, \citenamefont
  {Adolph}, \citenamefont {Aquila}, \citenamefont {Bajt}, \citenamefont
  {Barty}, \citenamefont {Bostedt}, \citenamefont {Bozek}, \citenamefont
  {Caleman}, \citenamefont {Coffee}, \citenamefont {Coppola}, \citenamefont
  {Delmas}, \citenamefont {Epp}, \citenamefont {Erk}, \citenamefont {Foucar},
  \citenamefont {Gorkhover}, \citenamefont {Gumprecht}, \citenamefont
  {Hartmann}, \citenamefont {Hartmann}, \citenamefont {Hauser}, \citenamefont
  {Holl}, \citenamefont {H{\"o}mke}, \citenamefont {Kimmel}, \citenamefont
  {Krasniqi}, \citenamefont {K{\"u}hnel}, \citenamefont {Maurer}, \citenamefont
  {Messerschmidt}, \citenamefont {Moshammer}, \citenamefont {Reich},
  \citenamefont {Rudek}, \citenamefont {Santra}, \citenamefont {Schlichting},
  \citenamefont {Schmidt}, \citenamefont {Schorb}, \citenamefont {Schulz},
  \citenamefont {Soltau}, \citenamefont {Spence}, \citenamefont {Starodub},
  \citenamefont {Str{\"u}der}, \citenamefont {Th{\o}gersen}, \citenamefont
  {Vrakking}, \citenamefont {Weidenspointner}, \citenamefont {White},
  \citenamefont {Wunderer}, \citenamefont {Meijer}, \citenamefont {Ullrich},
  \citenamefont {Stapelfeldt}, \citenamefont {Rolles},\ and\ \citenamefont
  {Chapman}}]{Kuepper:PRL112:083002}%
  \BibitemOpen
  \bibfield  {author} {\bibinfo {author} {\bibfnamefont {J.}~\bibnamefont
  {K{\"u}pper}}, \bibinfo {author} {\bibfnamefont {S.}~\bibnamefont {Stern}},
  \bibinfo {author} {\bibfnamefont {L.}~\bibnamefont {Holmegaard}}, \bibinfo
  {author} {\bibfnamefont {F.}~\bibnamefont {Filsinger}}, \bibinfo {author}
  {\bibfnamefont {A.}~\bibnamefont {Rouz\'{e}e}}, \bibinfo {author}
  {\bibfnamefont {A.}~\bibnamefont {Rudenko}}, \bibinfo {author} {\bibfnamefont
  {P.}~\bibnamefont {Johnsson}}, \bibinfo {author} {\bibfnamefont {A.~V.}\
  \bibnamefont {Martin}}, \bibinfo {author} {\bibfnamefont {M.}~\bibnamefont
  {Adolph}}, \bibinfo {author} {\bibfnamefont {A.}~\bibnamefont {Aquila}},
  \bibinfo {author} {\bibfnamefont {S.}~\bibnamefont {Bajt}}, \bibinfo {author}
  {\bibfnamefont {A.}~\bibnamefont {Barty}}, \bibinfo {author} {\bibfnamefont
  {C.}~\bibnamefont {Bostedt}}, \bibinfo {author} {\bibfnamefont
  {J.}~\bibnamefont {Bozek}}, \bibinfo {author} {\bibfnamefont
  {C.}~\bibnamefont {Caleman}}, \bibinfo {author} {\bibfnamefont
  {R.}~\bibnamefont {Coffee}}, \bibinfo {author} {\bibfnamefont
  {N.}~\bibnamefont {Coppola}}, \bibinfo {author} {\bibfnamefont
  {T.}~\bibnamefont {Delmas}}, \bibinfo {author} {\bibfnamefont
  {S.}~\bibnamefont {Epp}}, \bibinfo {author} {\bibfnamefont {B.}~\bibnamefont
  {Erk}}, \bibinfo {author} {\bibfnamefont {L.}~\bibnamefont {Foucar}},
  \bibinfo {author} {\bibfnamefont {T.}~\bibnamefont {Gorkhover}}, \bibinfo
  {author} {\bibfnamefont {L.}~\bibnamefont {Gumprecht}}, \bibinfo {author}
  {\bibfnamefont {A.}~\bibnamefont {Hartmann}}, \bibinfo {author}
  {\bibfnamefont {R.}~\bibnamefont {Hartmann}}, \bibinfo {author}
  {\bibfnamefont {G.}~\bibnamefont {Hauser}}, \bibinfo {author} {\bibfnamefont
  {P.}~\bibnamefont {Holl}}, \bibinfo {author} {\bibfnamefont {A.}~\bibnamefont
  {H{\"o}mke}}, \bibinfo {author} {\bibfnamefont {N.}~\bibnamefont {Kimmel}},
  \bibinfo {author} {\bibfnamefont {F.}~\bibnamefont {Krasniqi}}, \bibinfo
  {author} {\bibfnamefont {K.-U.}\ \bibnamefont {K{\"u}hnel}}, \bibinfo
  {author} {\bibfnamefont {J.}~\bibnamefont {Maurer}}, \bibinfo {author}
  {\bibfnamefont {M.}~\bibnamefont {Messerschmidt}}, \bibinfo {author}
  {\bibfnamefont {R.}~\bibnamefont {Moshammer}}, \bibinfo {author}
  {\bibfnamefont {C.}~\bibnamefont {Reich}}, \bibinfo {author} {\bibfnamefont
  {B.}~\bibnamefont {Rudek}}, \bibinfo {author} {\bibfnamefont
  {R.}~\bibnamefont {Santra}}, \bibinfo {author} {\bibfnamefont
  {I.}~\bibnamefont {Schlichting}}, \bibinfo {author} {\bibfnamefont
  {C.}~\bibnamefont {Schmidt}}, \bibinfo {author} {\bibfnamefont
  {S.}~\bibnamefont {Schorb}}, \bibinfo {author} {\bibfnamefont
  {J.}~\bibnamefont {Schulz}}, \bibinfo {author} {\bibfnamefont
  {H.}~\bibnamefont {Soltau}}, \bibinfo {author} {\bibfnamefont {J.~C.~H.}\
  \bibnamefont {Spence}}, \bibinfo {author} {\bibfnamefont {D.}~\bibnamefont
  {Starodub}}, \bibinfo {author} {\bibfnamefont {L.}~\bibnamefont
  {Str{\"u}der}}, \bibinfo {author} {\bibfnamefont {J.}~\bibnamefont
  {Th{\o}gersen}}, \bibinfo {author} {\bibfnamefont {M.~J.~J.}\ \bibnamefont
  {Vrakking}}, \bibinfo {author} {\bibfnamefont {G.}~\bibnamefont
  {Weidenspointner}}, \bibinfo {author} {\bibfnamefont {T.~A.}\ \bibnamefont
  {White}}, \bibinfo {author} {\bibfnamefont {C.}~\bibnamefont {Wunderer}},
  \bibinfo {author} {\bibfnamefont {G.}~\bibnamefont {Meijer}}, \bibinfo
  {author} {\bibfnamefont {J.}~\bibnamefont {Ullrich}}, \bibinfo {author}
  {\bibfnamefont {H.}~\bibnamefont {Stapelfeldt}}, \bibinfo {author}
  {\bibfnamefont {D.}~\bibnamefont {Rolles}}, \ and\ \bibinfo {author}
  {\bibfnamefont {H.~N.}\ \bibnamefont {Chapman}},\ }\bibfield  {title}
  {\enquote {\bibinfo {title} {X-ray diffraction from isolated and strongly
  aligned gas-phase molecules with a free-electron laser},}\ }\href {\doibase
  10.1103/PhysRevLett.112.083002} {\bibfield  {journal} {\bibinfo  {journal}
  {Phys.\ Rev.\ Lett.}\ }\textbf {\bibinfo {volume} {112}},\ \bibinfo {pages}
  {083002} (\bibinfo {year} {2014})},\ \Eprint {http://arxiv.org/abs/1307.4577}
  {arXiv:1307.4577 [physics]} \BibitemShut {NoStop}%
\bibitem [{\citenamefont {Yang}\ \emph {et~al.}(2016)\citenamefont {Yang},
  \citenamefont {Guehr}, \citenamefont {Vecchione}, \citenamefont {Robinson},
  \citenamefont {Li}, \citenamefont {Hartmann}, \citenamefont {Shen},
  \citenamefont {Coffee}, \citenamefont {Corbett}, \citenamefont {Fry},
  \citenamefont {Gaffney}, \citenamefont {Gorkhover}, \citenamefont {Hast},
  \citenamefont {Jobe}, \citenamefont {Makasyuk}, \citenamefont {Reid},
  \citenamefont {Robinson}, \citenamefont {Vetter}, \citenamefont {Wang},
  \citenamefont {Weathersby}, \citenamefont {Yoneda}, \citenamefont
  {Centurion},\ and\ \citenamefont {Wang}}]{Yang:NatComm7:11232}%
  \BibitemOpen
  \bibfield  {author} {\bibinfo {author} {\bibfnamefont {J.}~\bibnamefont
  {Yang}}, \bibinfo {author} {\bibfnamefont {M.}~\bibnamefont {Guehr}},
  \bibinfo {author} {\bibfnamefont {T.}~\bibnamefont {Vecchione}}, \bibinfo
  {author} {\bibfnamefont {M.~S.}\ \bibnamefont {Robinson}}, \bibinfo {author}
  {\bibfnamefont {R.}~\bibnamefont {Li}}, \bibinfo {author} {\bibfnamefont
  {N.}~\bibnamefont {Hartmann}}, \bibinfo {author} {\bibfnamefont
  {X.}~\bibnamefont {Shen}}, \bibinfo {author} {\bibfnamefont {R.}~\bibnamefont
  {Coffee}}, \bibinfo {author} {\bibfnamefont {J.}~\bibnamefont {Corbett}},
  \bibinfo {author} {\bibfnamefont {A.}~\bibnamefont {Fry}}, \bibinfo {author}
  {\bibfnamefont {K.}~\bibnamefont {Gaffney}}, \bibinfo {author} {\bibfnamefont
  {T.}~\bibnamefont {Gorkhover}}, \bibinfo {author} {\bibfnamefont
  {C.}~\bibnamefont {Hast}}, \bibinfo {author} {\bibfnamefont {K.}~\bibnamefont
  {Jobe}}, \bibinfo {author} {\bibfnamefont {I.}~\bibnamefont {Makasyuk}},
  \bibinfo {author} {\bibfnamefont {A.}~\bibnamefont {Reid}}, \bibinfo {author}
  {\bibfnamefont {J.}~\bibnamefont {Robinson}}, \bibinfo {author}
  {\bibfnamefont {S.}~\bibnamefont {Vetter}}, \bibinfo {author} {\bibfnamefont
  {F.}~\bibnamefont {Wang}}, \bibinfo {author} {\bibfnamefont {S.}~\bibnamefont
  {Weathersby}}, \bibinfo {author} {\bibfnamefont {C.}~\bibnamefont {Yoneda}},
  \bibinfo {author} {\bibfnamefont {M.}~\bibnamefont {Centurion}}, \ and\
  \bibinfo {author} {\bibfnamefont {X.}~\bibnamefont {Wang}},\ }\bibfield
  {title} {\enquote {\bibinfo {title} {Diffractive imaging of a rotational
  wavepacket in nitrogen molecules with femtosecond megaelectronvolt electron
  pulses},}\ }\href {\doibase 10.1038/ncomms11232} {\bibfield  {journal}
  {\bibinfo  {journal} {Nat. Commun.}\ }\textbf {\bibinfo {volume} {7}},\
  \bibinfo {pages} {11232} (\bibinfo {year} {2016})}\BibitemShut {NoStop}%
\bibitem [{\citenamefont {Brooks}(1976)}]{Brooks:Science193:11}%
  \BibitemOpen
  \bibfield  {author} {\bibinfo {author} {\bibfnamefont {P.~R.}\ \bibnamefont
  {Brooks}},\ }\bibfield  {title} {\enquote {\bibinfo {title} {Reactions of
  oriented molecules},}\ }\href {\doibase 10.1126/science.193.4247.11}
  {\bibfield  {journal} {\bibinfo  {journal} {Science}\ }\textbf {\bibinfo
  {volume} {193}},\ \bibinfo {pages} {11} (\bibinfo {year} {1976})}\BibitemShut
  {NoStop}%
\bibitem [{\citenamefont {Loesch}\ and\ \citenamefont
  {Remscheid}(1990)}]{Loesch:JCP93:4779}%
  \BibitemOpen
  \bibfield  {author} {\bibinfo {author} {\bibfnamefont {H.~J.}\ \bibnamefont
  {Loesch}}\ and\ \bibinfo {author} {\bibfnamefont {A.}~\bibnamefont
  {Remscheid}},\ }\bibfield  {title} {\enquote {\bibinfo {title} {Brute force
  in molecular reaction dynamics: {A} novel technique for measuring steric
  effects},}\ }\href {\doibase 10.1063/1.458668} {\bibfield  {journal}
  {\bibinfo  {journal} {J.\ Chem.\ Phys.}\ }\textbf {\bibinfo {volume} {93}},\
  \bibinfo {pages} {4779} (\bibinfo {year} {1990})}\BibitemShut {NoStop}%
\bibitem [{\citenamefont {Rakitzis}, \citenamefont {van~den Brom},\ and\
  \citenamefont {Janssen}(2004)}]{Rakitzis:Science303:1852}%
  \BibitemOpen
  \bibfield  {author} {\bibinfo {author} {\bibfnamefont {T.~P.}\ \bibnamefont
  {Rakitzis}}, \bibinfo {author} {\bibfnamefont {A.~J.}\ \bibnamefont {van~den
  Brom}}, \ and\ \bibinfo {author} {\bibfnamefont {M.~H.~M.}\ \bibnamefont
  {Janssen}},\ }\bibfield  {title} {\enquote {\bibinfo {title} {Directional
  dynamics in the photodissociation of oriented molecules},}\ }\href {\doibase
  10.1126/science.1094186} {\bibfield  {journal} {\bibinfo  {journal}
  {Science}\ }\textbf {\bibinfo {volume} {303}},\ \bibinfo {pages} {1852--1854}
  (\bibinfo {year} {2004})}\BibitemShut {NoStop}%
\bibitem [{\citenamefont {Stern}\ \emph {et~al.}(2014)\citenamefont {Stern},
  \citenamefont {Holmegaard}, \citenamefont {Filsinger}, \citenamefont
  {Rouzee}, \citenamefont {Rudenko}, \citenamefont {Johnsson}, \citenamefont
  {Martin}, \citenamefont {Barty}, \citenamefont {Bostedt}, \citenamefont
  {Bozek}, \citenamefont {Coffee}, \citenamefont {Epp}, \citenamefont {Erk},
  \citenamefont {Foucar}, \citenamefont {Hartmann}, \citenamefont {Kimmel},
  \citenamefont {K{\"u}hnel}, \citenamefont {Maurer}, \citenamefont
  {Messerschmidt}, \citenamefont {Rudek}, \citenamefont {Starodub},
  \citenamefont {Thøgersen}, \citenamefont {Weidenspointner}, \citenamefont
  {White}, \citenamefont {Stapelfeldt}, \citenamefont {Rolles}, \citenamefont
  {Chapman},\ and\ \citenamefont {K{\"u}pper}}]{Stern:FD171:393}%
  \BibitemOpen
  \bibfield  {author} {\bibinfo {author} {\bibfnamefont {S.}~\bibnamefont
  {Stern}}, \bibinfo {author} {\bibfnamefont {L.}~\bibnamefont {Holmegaard}},
  \bibinfo {author} {\bibfnamefont {F.}~\bibnamefont {Filsinger}}, \bibinfo
  {author} {\bibfnamefont {A.}~\bibnamefont {Rouzee}}, \bibinfo {author}
  {\bibfnamefont {A.}~\bibnamefont {Rudenko}}, \bibinfo {author} {\bibfnamefont
  {P.}~\bibnamefont {Johnsson}}, \bibinfo {author} {\bibfnamefont {A.~V.}\
  \bibnamefont {Martin}}, \bibinfo {author} {\bibfnamefont {A.}~\bibnamefont
  {Barty}}, \bibinfo {author} {\bibfnamefont {C.}~\bibnamefont {Bostedt}},
  \bibinfo {author} {\bibfnamefont {J.}~\bibnamefont {Bozek}}, \bibinfo
  {author} {\bibfnamefont {R.}~\bibnamefont {Coffee}}, \bibinfo {author}
  {\bibfnamefont {S.}~\bibnamefont {Epp}}, \bibinfo {author} {\bibfnamefont
  {B.}~\bibnamefont {Erk}}, \bibinfo {author} {\bibfnamefont {L.}~\bibnamefont
  {Foucar}}, \bibinfo {author} {\bibfnamefont {R.}~\bibnamefont {Hartmann}},
  \bibinfo {author} {\bibfnamefont {N.}~\bibnamefont {Kimmel}}, \bibinfo
  {author} {\bibfnamefont {K.-U.}\ \bibnamefont {K{\"u}hnel}}, \bibinfo
  {author} {\bibfnamefont {J.}~\bibnamefont {Maurer}}, \bibinfo {author}
  {\bibfnamefont {M.}~\bibnamefont {Messerschmidt}}, \bibinfo {author}
  {\bibfnamefont {B.}~\bibnamefont {Rudek}}, \bibinfo {author} {\bibfnamefont
  {D.}~\bibnamefont {Starodub}}, \bibinfo {author} {\bibfnamefont
  {J.}~\bibnamefont {Thøgersen}}, \bibinfo {author} {\bibfnamefont
  {G.}~\bibnamefont {Weidenspointner}}, \bibinfo {author} {\bibfnamefont
  {T.~A.}\ \bibnamefont {White}}, \bibinfo {author} {\bibfnamefont
  {H.}~\bibnamefont {Stapelfeldt}}, \bibinfo {author} {\bibfnamefont
  {D.}~\bibnamefont {Rolles}}, \bibinfo {author} {\bibfnamefont {H.~N.}\
  \bibnamefont {Chapman}}, \ and\ \bibinfo {author} {\bibfnamefont
  {J.}~\bibnamefont {K{\"u}pper}},\ }\bibfield  {title} {\enquote {\bibinfo
  {title} {Toward atomic resolution diffractive imaging of isolated molecules
  with x-ray free-electron lasers},}\ }\href {\doibase 10.1039/c4fd00028e}
  {\bibfield  {journal} {\bibinfo  {journal} {Faraday Disc.}\ }\textbf
  {\bibinfo {volume} {171}},\ \bibinfo {pages} {393} (\bibinfo {year}
  {2014})},\ \Eprint {http://arxiv.org/abs/1403.2553} {arXiv:1403.2553
  [physics]} \BibitemShut {NoStop}%
\bibitem [{\citenamefont {Filsinger}\ \emph {et~al.}(2011)\citenamefont
  {Filsinger}, \citenamefont {Meijer}, \citenamefont {Stapelfeldt},
  \citenamefont {Chapman},\ and\ \citenamefont
  {K{\"u}pper}}]{Filsinger:PCCP13:2076}%
  \BibitemOpen
  \bibfield  {author} {\bibinfo {author} {\bibfnamefont {F.}~\bibnamefont
  {Filsinger}}, \bibinfo {author} {\bibfnamefont {G.}~\bibnamefont {Meijer}},
  \bibinfo {author} {\bibfnamefont {H.}~\bibnamefont {Stapelfeldt}}, \bibinfo
  {author} {\bibfnamefont {H.}~\bibnamefont {Chapman}}, \ and\ \bibinfo
  {author} {\bibfnamefont {J.}~\bibnamefont {K{\"u}pper}},\ }\bibfield  {title}
  {\enquote {\bibinfo {title} {State- and conformer-selected beams of aligned
  and oriented molecules for ultrafast diffraction studies},}\ }\href {\doibase
  10.1039/C0CP01585G} {\bibfield  {journal} {\bibinfo  {journal} {Phys.\ Chem.\
  Chem.\ Phys.}\ }\textbf {\bibinfo {volume} {13}},\ \bibinfo {pages}
  {2076--2087} (\bibinfo {year} {2011})}\BibitemShut {NoStop}%
\bibitem [{\citenamefont {Spence}\ and\ \citenamefont
  {Doak}(2004)}]{Spence:PRL92:198102}%
  \BibitemOpen
  \bibfield  {author} {\bibinfo {author} {\bibfnamefont {J.~C.~H.}\
  \bibnamefont {Spence}}\ and\ \bibinfo {author} {\bibfnamefont {R.~B.}\
  \bibnamefont {Doak}},\ }\bibfield  {title} {\enquote {\bibinfo {title}
  {Single molecule diffraction},}\ }\href {\doibase
  10.1103/PhysRevLett.92.198102} {\bibfield  {journal} {\bibinfo  {journal}
  {Phys.\ Rev.\ Lett.}\ }\textbf {\bibinfo {volume} {92}},\ \bibinfo {pages}
  {198102} (\bibinfo {year} {2004})}\BibitemShut {NoStop}%
\bibitem [{\citenamefont {Friedrich}\ and\ \citenamefont
  {Herschbach}(1991)}]{Friedrich:Nature353:412}%
  \BibitemOpen
  \bibfield  {author} {\bibinfo {author} {\bibfnamefont {B.}~\bibnamefont
  {Friedrich}}\ and\ \bibinfo {author} {\bibfnamefont {D.~R.}\ \bibnamefont
  {Herschbach}},\ }\bibfield  {title} {\enquote {\bibinfo {title} {Spatial
  orientation of molecules in strong electric fields and evidence for pendular
  states},}\ }\href {\doibase 10.1038/353412a0} {\bibfield  {journal} {\bibinfo
   {journal} {Nature}\ }\textbf {\bibinfo {volume} {353}},\ \bibinfo {pages}
  {412--414} (\bibinfo {year} {1991})}\BibitemShut {NoStop}%
\bibitem [{\citenamefont {Block}, \citenamefont {Bohac},\ and\ \citenamefont
  {Miller}(1992)}]{Block:PRL68:1303}%
  \BibitemOpen
  \bibfield  {author} {\bibinfo {author} {\bibfnamefont {P.~A.}\ \bibnamefont
  {Block}}, \bibinfo {author} {\bibfnamefont {E.~J.}\ \bibnamefont {Bohac}}, \
  and\ \bibinfo {author} {\bibfnamefont {R.~E.}\ \bibnamefont {Miller}},\
  }\bibfield  {title} {\enquote {\bibinfo {title} {Spectroscopy of pendular
  states -- the use of molecular complexes in achieving orientation},}\ }\href
  {\doibase 10.1103/PhysRevLett.68.1303} {\bibfield  {journal} {\bibinfo
  {journal} {Phys.\ Rev.\ Lett.}\ }\textbf {\bibinfo {volume} {68}},\ \bibinfo
  {pages} {1303--1306} (\bibinfo {year} {1992})}\BibitemShut {NoStop}%
\bibitem [{\citenamefont {Slenczka}, \citenamefont {Friedrich},\ and\
  \citenamefont {Herschbach}(1994)}]{Slenczka:PRL72:1806}%
  \BibitemOpen
  \bibfield  {author} {\bibinfo {author} {\bibfnamefont {A.}~\bibnamefont
  {Slenczka}}, \bibinfo {author} {\bibfnamefont {B.}~\bibnamefont {Friedrich}},
  \ and\ \bibinfo {author} {\bibfnamefont {D.}~\bibnamefont {Herschbach}},\
  }\bibfield  {title} {\enquote {\bibinfo {title} {Pendular alignment of
  paramagnetic molecules in uniform magnetic fields},}\ }\href {\doibase
  10.1103/PhysRevLett.72.1806} {\bibfield  {journal} {\bibinfo  {journal}
  {Phys.\ Rev.\ Lett.}\ }\textbf {\bibinfo {volume} {72}},\ \bibinfo {pages}
  {1806--1809} (\bibinfo {year} {1994})}\BibitemShut {NoStop}%
\bibitem [{\citenamefont {Ghafur}\ \emph {et~al.}(2009)\citenamefont {Ghafur},
  \citenamefont {Rouzee}, \citenamefont {Gijsbertsen}, \citenamefont {Siu},
  \citenamefont {Stolte},\ and\ \citenamefont
  {Vrakking}}]{Ghafur:NatPhys5:289}%
  \BibitemOpen
  \bibfield  {author} {\bibinfo {author} {\bibfnamefont {O.}~\bibnamefont
  {Ghafur}}, \bibinfo {author} {\bibfnamefont {A.}~\bibnamefont {Rouzee}},
  \bibinfo {author} {\bibfnamefont {A.}~\bibnamefont {Gijsbertsen}}, \bibinfo
  {author} {\bibfnamefont {W.~K.}\ \bibnamefont {Siu}}, \bibinfo {author}
  {\bibfnamefont {S.}~\bibnamefont {Stolte}}, \ and\ \bibinfo {author}
  {\bibfnamefont {M.~J.~J.}\ \bibnamefont {Vrakking}},\ }\bibfield  {title}
  {\enquote {\bibinfo {title} {Impulsive orientation and alignment of
  quantum-state-selected {NO} molecules},}\ }\href {\doibase 10.1038/nphys1225}
  {\bibfield  {journal} {\bibinfo  {journal} {Nat. Phys.}\ }\textbf {\bibinfo
  {volume} {5}},\ \bibinfo {pages} {289--293} (\bibinfo {year}
  {2009})}\BibitemShut {NoStop}%
\bibitem [{\citenamefont {Goban}, \citenamefont {Minemoto},\ and\ \citenamefont
  {Sakai}(2008)}]{Goban:PRL101:013001}%
  \BibitemOpen
  \bibfield  {author} {\bibinfo {author} {\bibfnamefont {A.}~\bibnamefont
  {Goban}}, \bibinfo {author} {\bibfnamefont {S.}~\bibnamefont {Minemoto}}, \
  and\ \bibinfo {author} {\bibfnamefont {H.}~\bibnamefont {Sakai}},\ }\bibfield
   {title} {\enquote {\bibinfo {title} {Laser-field-free molecular
  orientation},}\ }\href {\doibase 10.1103/PhysRevLett.101.013001} {\bibfield
  {journal} {\bibinfo  {journal} {Phys.\ Rev.\ Lett.}\ }\textbf {\bibinfo
  {volume} {101}},\ \bibinfo {pages} {013001} (\bibinfo {year}
  {2008})}\BibitemShut {NoStop}%
\bibitem [{\citenamefont {Vrakking}\ and\ \citenamefont
  {Stolte}(1997)}]{Vrakking:CPL271:209}%
  \BibitemOpen
  \bibfield  {author} {\bibinfo {author} {\bibfnamefont {M.~J.~J.}\
  \bibnamefont {Vrakking}}\ and\ \bibinfo {author} {\bibfnamefont
  {S.}~\bibnamefont {Stolte}},\ }\bibfield  {title} {\enquote {\bibinfo {title}
  {Coherent control of molecular orientation},}\ }\href {\doibase
  10.1016/S0009-2614(97)00436-3} {\bibfield  {journal} {\bibinfo  {journal}
  {Chem.\ Phys.\ Lett.}\ }\textbf {\bibinfo {volume} {271}},\ \bibinfo {pages}
  {209--215} (\bibinfo {year} {1997})}\BibitemShut {NoStop}%
\bibitem [{\citenamefont {De}\ \emph {et~al.}(2009)\citenamefont {De},
  \citenamefont {Znakovskaya}, \citenamefont {Ray}, \citenamefont {Anis},
  \citenamefont {Johnson}, \citenamefont {Bocharova}, \citenamefont
  {Magrakvelidze}, \citenamefont {Esry}, \citenamefont {Cocke}, \citenamefont
  {Litvinyuk},\ and\ \citenamefont {Kling}}]{De:PRL103:153002}%
  \BibitemOpen
  \bibfield  {author} {\bibinfo {author} {\bibfnamefont {S.}~\bibnamefont
  {De}}, \bibinfo {author} {\bibfnamefont {I.}~\bibnamefont {Znakovskaya}},
  \bibinfo {author} {\bibfnamefont {D.}~\bibnamefont {Ray}}, \bibinfo {author}
  {\bibfnamefont {F.}~\bibnamefont {Anis}}, \bibinfo {author} {\bibfnamefont
  {N.~G.}\ \bibnamefont {Johnson}}, \bibinfo {author} {\bibfnamefont {I.~A.}\
  \bibnamefont {Bocharova}}, \bibinfo {author} {\bibfnamefont {M.}~\bibnamefont
  {Magrakvelidze}}, \bibinfo {author} {\bibfnamefont {B.~D.}\ \bibnamefont
  {Esry}}, \bibinfo {author} {\bibfnamefont {C.~L.}\ \bibnamefont {Cocke}},
  \bibinfo {author} {\bibfnamefont {I.~V.}\ \bibnamefont {Litvinyuk}}, \ and\
  \bibinfo {author} {\bibfnamefont {M.~F.}\ \bibnamefont {Kling}},\ }\bibfield
  {title} {\enquote {\bibinfo {title} {Field-free orientation of {CO} molecules
  by femtosecond two-color laser fields},}\ }\href {\doibase
  10.1103/PhysRevLett.103.153002} {\bibfield  {journal} {\bibinfo  {journal}
  {Phys.\ Rev.\ Lett.}\ }\textbf {\bibinfo {volume} {103}},\ \bibinfo {pages}
  {153002} (\bibinfo {year} {2009})},\ \Eprint {http://arxiv.org/abs/0907.3250}
  {arXiv:0907.3250 [physics.chem-ph]} \BibitemShut {NoStop}%
\bibitem [{\citenamefont {Harde}, \citenamefont {Keiding},\ and\ \citenamefont
  {Grischkowsky}(1991)}]{Harde:PRL66:1834}%
  \BibitemOpen
  \bibfield  {author} {\bibinfo {author} {\bibfnamefont {H.}~\bibnamefont
  {Harde}}, \bibinfo {author} {\bibfnamefont {S.}~\bibnamefont {Keiding}}, \
  and\ \bibinfo {author} {\bibfnamefont {D.}~\bibnamefont {Grischkowsky}},\
  }\bibfield  {title} {\enquote {\bibinfo {title} {{THZ} commensurate echoes -
  periodic rephasing of molecular-transitions in free-induction decay},}\
  }\href@noop {} {\bibfield  {journal} {\bibinfo  {journal} {Phys.\ Rev.\
  Lett.}\ }\textbf {\bibinfo {volume} {66}},\ \bibinfo {pages} {1834--1837}
  (\bibinfo {year} {1991})}\BibitemShut {NoStop}%
\bibitem [{\citenamefont {Machholm}\ and\ \citenamefont
  {Henriksen}(2001)}]{Machholm:PRL87:193001}%
  \BibitemOpen
  \bibfield  {author} {\bibinfo {author} {\bibfnamefont {M.}~\bibnamefont
  {Machholm}}\ and\ \bibinfo {author} {\bibfnamefont {N.~E.}\ \bibnamefont
  {Henriksen}},\ }\bibfield  {title} {\enquote {\bibinfo {title} {Field-free
  orientation of molecules},}\ }\href {\doibase 10.1103/PhysRevLett.87.193001}
  {\bibfield  {journal} {\bibinfo  {journal} {Phys.\ Rev.\ Lett.}\ }\textbf
  {\bibinfo {volume} {87}},\ \bibinfo {pages} {193001} (\bibinfo {year}
  {2001})}\BibitemShut {NoStop}%
\bibitem [{\citenamefont {Fleischer}\ \emph {et~al.}(2011)\citenamefont
  {Fleischer}, \citenamefont {Zhou}, \citenamefont {Field},\ and\ \citenamefont
  {Nelson}}]{Fleischer:PRL107:163603}%
  \BibitemOpen
  \bibfield  {author} {\bibinfo {author} {\bibfnamefont {S.}~\bibnamefont
  {Fleischer}}, \bibinfo {author} {\bibfnamefont {Y.}~\bibnamefont {Zhou}},
  \bibinfo {author} {\bibfnamefont {R.~W.}\ \bibnamefont {Field}}, \ and\
  \bibinfo {author} {\bibfnamefont {K.~A.}\ \bibnamefont {Nelson}},\ }\bibfield
   {title} {\enquote {\bibinfo {title} {Molecular orientation and alignment by
  intense single-cycle {THz} pulses},}\ }\href {\doibase
  10.1103/PhysRevLett.107.163603} {\bibfield  {journal} {\bibinfo  {journal}
  {Phys.\ Rev.\ Lett.}\ }\textbf {\bibinfo {volume} {107}},\ \bibinfo {pages}
  {163603} (\bibinfo {year} {2011})},\ \Eprint {http://arxiv.org/abs/1105.1635}
  {arXiv:1105.1635 [physics.chem-ph]} \BibitemShut {NoStop}%
\bibitem [{\citenamefont {Egodapitiya}, \citenamefont {Li},\ and\ \citenamefont
  {Jones}(2014)}]{Egodapitiya:PRL112:103002}%
  \BibitemOpen
  \bibfield  {author} {\bibinfo {author} {\bibfnamefont {K.~N.}\ \bibnamefont
  {Egodapitiya}}, \bibinfo {author} {\bibfnamefont {S.}~\bibnamefont {Li}}, \
  and\ \bibinfo {author} {\bibfnamefont {R.~R.}\ \bibnamefont {Jones}},\
  }\bibfield  {title} {\enquote {\bibinfo {title} {Terahertz-induced field-free
  orientation of rotationally excited molecules},}\ }\href {\doibase
  10.1103/PhysRevLett.112.103002} {\bibfield  {journal} {\bibinfo  {journal}
  {Phys.\ Rev.\ Lett.}\ }\textbf {\bibinfo {volume} {112}},\ \bibinfo {pages}
  {103002} (\bibinfo {year} {2014})}\BibitemShut {NoStop}%
\bibitem [{\citenamefont {Friedrich}\ and\ \citenamefont
  {Herschbach}(1999)}]{Friedrich:JCP111:6157}%
  \BibitemOpen
  \bibfield  {author} {\bibinfo {author} {\bibfnamefont {B.}~\bibnamefont
  {Friedrich}}\ and\ \bibinfo {author} {\bibfnamefont {D.}~\bibnamefont
  {Herschbach}},\ }\bibfield  {title} {\enquote {\bibinfo {title} {Enhanced
  orientation of polar molecules by combined electrostatic and nonresonant
  induced dipole forces},}\ }\href {\doibase 10.1063/1.479917} {\bibfield
  {journal} {\bibinfo  {journal} {J.\ Chem.\ Phys.}\ }\textbf {\bibinfo
  {volume} {111}},\ \bibinfo {pages} {6157} (\bibinfo {year}
  {1999})}\BibitemShut {NoStop}%
\bibitem [{\citenamefont {Holmegaard}\ \emph {et~al.}(2009)\citenamefont
  {Holmegaard}, \citenamefont {Nielsen}, \citenamefont {Nevo}, \citenamefont
  {Stapelfeldt}, \citenamefont {Filsinger}, \citenamefont {K{\"u}pper},\ and\
  \citenamefont {Meijer}}]{Holmegaard:PRL102:023001}%
  \BibitemOpen
  \bibfield  {author} {\bibinfo {author} {\bibfnamefont {L.}~\bibnamefont
  {Holmegaard}}, \bibinfo {author} {\bibfnamefont {J.~H.}\ \bibnamefont
  {Nielsen}}, \bibinfo {author} {\bibfnamefont {I.}~\bibnamefont {Nevo}},
  \bibinfo {author} {\bibfnamefont {H.}~\bibnamefont {Stapelfeldt}}, \bibinfo
  {author} {\bibfnamefont {F.}~\bibnamefont {Filsinger}}, \bibinfo {author}
  {\bibfnamefont {J.}~\bibnamefont {K{\"u}pper}}, \ and\ \bibinfo {author}
  {\bibfnamefont {G.}~\bibnamefont {Meijer}},\ }\bibfield  {title} {\enquote
  {\bibinfo {title} {Laser-induced alignment and orientation of
  quantum-state-selected large molecules},}\ }\href {\doibase
  10.1103/PhysRevLett.102.023001} {\bibfield  {journal} {\bibinfo  {journal}
  {Phys.\ Rev.\ Lett.}\ }\textbf {\bibinfo {volume} {102}},\ \bibinfo {pages}
  {023001} (\bibinfo {year} {2009})},\ \Eprint {http://arxiv.org/abs/0810.2307}
  {arXiv:0810.2307 [physics.chem-ph]} \BibitemShut {NoStop}%
\bibitem [{\citenamefont {Filsinger}\ \emph
  {et~al.}(2009{\natexlab{a}})\citenamefont {Filsinger}, \citenamefont
  {K{\"u}pper}, \citenamefont {Meijer}, \citenamefont {Holmegaard},
  \citenamefont {Nielsen}, \citenamefont {Nevo}, \citenamefont {Hansen},\ and\
  \citenamefont {Stapelfeldt}}]{Filsinger:JCP131:064309}%
  \BibitemOpen
  \bibfield  {author} {\bibinfo {author} {\bibfnamefont {F.}~\bibnamefont
  {Filsinger}}, \bibinfo {author} {\bibfnamefont {J.}~\bibnamefont
  {K{\"u}pper}}, \bibinfo {author} {\bibfnamefont {G.}~\bibnamefont {Meijer}},
  \bibinfo {author} {\bibfnamefont {L.}~\bibnamefont {Holmegaard}}, \bibinfo
  {author} {\bibfnamefont {J.~H.}\ \bibnamefont {Nielsen}}, \bibinfo {author}
  {\bibfnamefont {I.}~\bibnamefont {Nevo}}, \bibinfo {author} {\bibfnamefont
  {J.~L.}\ \bibnamefont {Hansen}}, \ and\ \bibinfo {author} {\bibfnamefont
  {H.}~\bibnamefont {Stapelfeldt}},\ }\bibfield  {title} {\enquote {\bibinfo
  {title} {Quantum-state selection, alignment, and orientation of large
  molecules using static electric and laser fields},}\ }\href {\doibase
  10.1063/1.3194287} {\bibfield  {journal} {\bibinfo  {journal} {J.\ Chem.\
  Phys.}\ }\textbf {\bibinfo {volume} {131}},\ \bibinfo {pages} {064309}
  (\bibinfo {year} {2009}{\natexlab{a}})},\ \Eprint
  {http://arxiv.org/abs/0903.5413} {arXiv:0903.5413 [physics]} \BibitemShut
  {NoStop}%
\bibitem [{\citenamefont {Trippel}\ \emph {et~al.}(2013)\citenamefont
  {Trippel}, \citenamefont {Mullins}, \citenamefont {M{\"u}ller}, \citenamefont
  {Kienitz}, \citenamefont {D{\l}ugo{\l}\k{e}cki},\ and\ \citenamefont
  {K{\"u}pper}}]{Trippel:MP111:1738}%
  \BibitemOpen
  \bibfield  {author} {\bibinfo {author} {\bibfnamefont {S.}~\bibnamefont
  {Trippel}}, \bibinfo {author} {\bibfnamefont {T.}~\bibnamefont {Mullins}},
  \bibinfo {author} {\bibfnamefont {N.~L.~M.}\ \bibnamefont {M{\"u}ller}},
  \bibinfo {author} {\bibfnamefont {J.~S.}\ \bibnamefont {Kienitz}}, \bibinfo
  {author} {\bibfnamefont {K.}~\bibnamefont {D{\l}ugo{\l}\k{e}cki}}, \ and\
  \bibinfo {author} {\bibfnamefont {J.}~\bibnamefont {K{\"u}pper}},\ }\bibfield
   {title} {\enquote {\bibinfo {title} {Strongly aligned and oriented molecular
  samples at a {kHz} repetition rate},}\ }\href {\doibase
  10.1080/00268976.2013.780334} {\bibfield  {journal} {\bibinfo  {journal}
  {Mol.\ Phys.}\ }\textbf {\bibinfo {volume} {111}},\ \bibinfo {pages} {1738}
  (\bibinfo {year} {2013})},\ \Eprint {http://arxiv.org/abs/1301.1826}
  {arXiv:1301.1826 [physics.atom-ph]} \BibitemShut {NoStop}%
\bibitem [{\citenamefont {Stern}(1926)}]{Stern:ZP39:751}%
  \BibitemOpen
  \bibfield  {author} {\bibinfo {author} {\bibfnamefont {O.}~\bibnamefont
  {Stern}},\ }\bibfield  {title} {\enquote {\bibinfo {title} {{Z}ur {M}ethode
  der {M}olekularstrahlen {I}},}\ }\href {\doibase 10.1007/BF01451746}
  {\bibfield  {journal} {\bibinfo  {journal} {Z.\ Phys.}\ }\textbf {\bibinfo
  {volume} {39}},\ \bibinfo {pages} {751--763} (\bibinfo {year}
  {1926})}\BibitemShut {NoStop}%
\bibitem [{\citenamefont {Reuss}(1988)}]{Reuss:StateSelection}%
  \BibitemOpen
  \bibfield  {author} {\bibinfo {author} {\bibfnamefont {J.}~\bibnamefont
  {Reuss}},\ }\bibfield  {title} {\enquote {\bibinfo {title} {{S}tate
  {S}election by {N}onoptical {M}ethods},}\ }in\ \href@noop {} {\emph {\bibinfo
  {booktitle} {Atomic and molecular beam methods}}},\ Vol.~\bibinfo {volume}
  {1},\ \bibinfo {editor} {edited by\ \bibinfo {editor} {\bibfnamefont
  {G.}~\bibnamefont {Scoles}}}\ (\bibinfo  {publisher} {Oxford University
  Press},\ \bibinfo {address} {New York, NY, USA},\ \bibinfo {year} {1988})\
  Chap.~\bibinfo {chapter} {11}, pp.\ \bibinfo {pages} {276--292}\BibitemShut
  {NoStop}%
\bibitem [{\citenamefont {Nielsen}\ \emph {et~al.}(2011)\citenamefont
  {Nielsen}, \citenamefont {Simesen}, \citenamefont {Bisgaard}, \citenamefont
  {Stapelfeldt}, \citenamefont {Filsinger}, \citenamefont {Friedrich},
  \citenamefont {Meijer},\ and\ \citenamefont
  {K{\"u}pper}}]{Nielsen:PCCP13:18971}%
  \BibitemOpen
  \bibfield  {author} {\bibinfo {author} {\bibfnamefont {J.~H.}\ \bibnamefont
  {Nielsen}}, \bibinfo {author} {\bibfnamefont {P.}~\bibnamefont {Simesen}},
  \bibinfo {author} {\bibfnamefont {C.~Z.}\ \bibnamefont {Bisgaard}}, \bibinfo
  {author} {\bibfnamefont {H.}~\bibnamefont {Stapelfeldt}}, \bibinfo {author}
  {\bibfnamefont {F.}~\bibnamefont {Filsinger}}, \bibinfo {author}
  {\bibfnamefont {B.}~\bibnamefont {Friedrich}}, \bibinfo {author}
  {\bibfnamefont {G.}~\bibnamefont {Meijer}}, \ and\ \bibinfo {author}
  {\bibfnamefont {J.}~\bibnamefont {K{\"u}pper}},\ }\bibfield  {title}
  {\enquote {\bibinfo {title} {Stark-selected beam of ground-state {OCS}
  molecules characterized by revivals of impulsive alignment},}\ }\href
  {\doibase 10.1039/c1cp21143a} {\bibfield  {journal} {\bibinfo  {journal}
  {Phys.\ Chem.\ Chem.\ Phys.}\ }\textbf {\bibinfo {volume} {13}},\ \bibinfo
  {pages} {18971--18975} (\bibinfo {year} {2011})},\ \Eprint
  {http://arxiv.org/abs/1105.2413} {arXiv:1105.2413 [physics]} \BibitemShut
  {NoStop}%
\bibitem [{\citenamefont {Putzke}\ \emph {et~al.}(2011)\citenamefont {Putzke},
  \citenamefont {Filsinger}, \citenamefont {Haak}, \citenamefont {K{\"u}pper},\
  and\ \citenamefont {Meijer}}]{Putzke:PCCP13:18962}%
  \BibitemOpen
  \bibfield  {author} {\bibinfo {author} {\bibfnamefont {S.}~\bibnamefont
  {Putzke}}, \bibinfo {author} {\bibfnamefont {F.}~\bibnamefont {Filsinger}},
  \bibinfo {author} {\bibfnamefont {H.}~\bibnamefont {Haak}}, \bibinfo {author}
  {\bibfnamefont {J.}~\bibnamefont {K{\"u}pper}}, \ and\ \bibinfo {author}
  {\bibfnamefont {G.}~\bibnamefont {Meijer}},\ }\bibfield  {title} {\enquote
  {\bibinfo {title} {Rotational-state-specific guiding of large molecules},}\
  }\href {\doibase 10.1039/C1CP20721K} {\bibfield  {journal} {\bibinfo
  {journal} {Phys.\ Chem.\ Chem.\ Phys.}\ }\textbf {\bibinfo {volume} {13}},\
  \bibinfo {pages} {18962} (\bibinfo {year} {2011})},\ \Eprint
  {http://arxiv.org/abs/1103.5080} {arXiv:1103.5080 [physics]} \BibitemShut
  {NoStop}%
\bibitem [{\citenamefont {Chang}\ \emph {et~al.}(2015)\citenamefont {Chang},
  \citenamefont {Horke}, \citenamefont {Trippel},\ and\ \citenamefont
  {K{\"u}pper}}]{Chang:IRPC34:557}%
  \BibitemOpen
  \bibfield  {author} {\bibinfo {author} {\bibfnamefont {Y.-P.}\ \bibnamefont
  {Chang}}, \bibinfo {author} {\bibfnamefont {D.}~\bibnamefont {Horke}},
  \bibinfo {author} {\bibfnamefont {S.}~\bibnamefont {Trippel}}, \ and\
  \bibinfo {author} {\bibfnamefont {J.}~\bibnamefont {K{\"u}pper}},\ }\bibfield
   {title} {\enquote {\bibinfo {title} {Spatially-controlled complex molecules
  and their applications},}\ }\href {\doibase 10.1080/0144235X.2015.1077838}
  {\bibfield  {journal} {\bibinfo  {journal} {Int.\ Rev.\ Phys.\ Chem.}\
  }\textbf {\bibinfo {volume} {34}},\ \bibinfo {pages} {557--590} (\bibinfo
  {year} {2015})},\ \Eprint {http://arxiv.org/abs/1505.05632} {arXiv:1505.05632
  [physics]} \BibitemShut {NoStop}%
\bibitem [{\citenamefont {Born}\ and\ \citenamefont
  {Fock}(1928)}]{Born:ZP51:165}%
  \BibitemOpen
  \bibfield  {author} {\bibinfo {author} {\bibfnamefont {M.}~\bibnamefont
  {Born}}\ and\ \bibinfo {author} {\bibfnamefont {V.}~\bibnamefont {Fock}},\
  }\bibfield  {title} {\enquote {\bibinfo {title} {{B}eweis des
  {A}diabatensatzes},}\ }\href {\doibase 10.1007/BF01343193} {\bibfield
  {journal} {\bibinfo  {journal} {Z.\ Phys.}\ }\textbf {\bibinfo {volume}
  {51}},\ \bibinfo {pages} {165--180} (\bibinfo {year} {1928})}\BibitemShut
  {NoStop}%
\bibitem [{\citenamefont {Ortigoso}(2012)}]{Ortigoso:PRA86:032121}%
  \BibitemOpen
  \bibfield  {author} {\bibinfo {author} {\bibfnamefont {J.}~\bibnamefont
  {Ortigoso}},\ }\bibfield  {title} {\enquote {\bibinfo {title} {Quantum
  adiabatic theorem in light of the {M}arzlin-{S}anders inconsistency},}\
  }\href {\doibase 10.1103/PhysRevA.86.032121} {\bibfield  {journal} {\bibinfo
  {journal} {Phys.\ Rev.\ A}\ }\textbf {\bibinfo {volume} {86}},\ \bibinfo
  {pages} {032121} (\bibinfo {year} {2012})},\ \Eprint
  {http://arxiv.org/abs/1111.5195} {arXiv:1111.5195 [quant-ph]} \BibitemShut
  {NoStop}%
\bibitem [{\citenamefont {Nielsen}\ \emph {et~al.}(2012)\citenamefont
  {Nielsen}, \citenamefont {Stapelfeldt}, \citenamefont {K{\"u}pper},
  \citenamefont {Friedrich}, \citenamefont {Omiste},\ and\ \citenamefont
  {Gonz{\'a}lez-F{\'e}rez}}]{Nielsen:PRL108:193001}%
  \BibitemOpen
  \bibfield  {author} {\bibinfo {author} {\bibfnamefont {J.~H.}\ \bibnamefont
  {Nielsen}}, \bibinfo {author} {\bibfnamefont {H.}~\bibnamefont
  {Stapelfeldt}}, \bibinfo {author} {\bibfnamefont {J.}~\bibnamefont
  {K{\"u}pper}}, \bibinfo {author} {\bibfnamefont {B.}~\bibnamefont
  {Friedrich}}, \bibinfo {author} {\bibfnamefont {J.~J.}\ \bibnamefont
  {Omiste}}, \ and\ \bibinfo {author} {\bibfnamefont {R.}~\bibnamefont
  {Gonz{\'a}lez-F{\'e}rez}},\ }\bibfield  {title} {\enquote {\bibinfo {title}
  {Making the best of mixed-field orientation of polar molecules: A recipe for
  achieving adiabatic dynamics in an electrostatic field combined with laser
  pulses},}\ }\href {\doibase 10.1103/PhysRevLett.108.193001} {\bibfield
  {journal} {\bibinfo  {journal} {Phys.\ Rev.\ Lett.}\ }\textbf {\bibinfo
  {volume} {108}},\ \bibinfo {pages} {193001} (\bibinfo {year} {2012})},\
  \Eprint {http://arxiv.org/abs/1204.2685} {arXiv:1204.2685 [physics.chem-ph]}
  \BibitemShut {NoStop}%
\bibitem [{\citenamefont {Trippel}\ \emph {et~al.}(2015)\citenamefont
  {Trippel}, \citenamefont {Mullins}, \citenamefont {M{\"u}ller}, \citenamefont
  {Kienitz}, \citenamefont {Gonz{\'a}lez-F{\'e}rez},\ and\ \citenamefont
  {K{\"u}pper}}]{Trippel:PRL114:103003}%
  \BibitemOpen
  \bibfield  {author} {\bibinfo {author} {\bibfnamefont {S.}~\bibnamefont
  {Trippel}}, \bibinfo {author} {\bibfnamefont {T.}~\bibnamefont {Mullins}},
  \bibinfo {author} {\bibfnamefont {N.~L.~M.}\ \bibnamefont {M{\"u}ller}},
  \bibinfo {author} {\bibfnamefont {J.~S.}\ \bibnamefont {Kienitz}}, \bibinfo
  {author} {\bibfnamefont {R.}~\bibnamefont {Gonz{\'a}lez-F{\'e}rez}}, \ and\
  \bibinfo {author} {\bibfnamefont {J.}~\bibnamefont {K{\"u}pper}},\ }\bibfield
   {title} {\enquote {\bibinfo {title} {Two-state wave packet for strong
  field-free molecular orientation},}\ }\href {\doibase
  10.1103/PhysRevLett.114.103003} {\bibfield  {journal} {\bibinfo  {journal}
  {Phys.\ Rev.\ Lett.}\ }\textbf {\bibinfo {volume} {114}},\ \bibinfo {pages}
  {103003} (\bibinfo {year} {2015})},\ \Eprint {http://arxiv.org/abs/1409.2836}
  {arXiv:1409.2836 [physics]} \BibitemShut {NoStop}%
\bibitem [{\citenamefont {Even}\ \emph {et~al.}(2000)\citenamefont {Even},
  \citenamefont {Jortner}, \citenamefont {Noy}, \citenamefont {Lavie},\ and\
  \citenamefont {Cossart-Magos}}]{Even:JCP112:8068}%
  \BibitemOpen
  \bibfield  {author} {\bibinfo {author} {\bibfnamefont {U.}~\bibnamefont
  {Even}}, \bibinfo {author} {\bibfnamefont {J.}~\bibnamefont {Jortner}},
  \bibinfo {author} {\bibfnamefont {D.}~\bibnamefont {Noy}}, \bibinfo {author}
  {\bibfnamefont {N.}~\bibnamefont {Lavie}}, \ and\ \bibinfo {author}
  {\bibfnamefont {N.}~\bibnamefont {Cossart-Magos}},\ }\bibfield  {title}
  {\enquote {\bibinfo {title} {Cooling of large molecules below 1~{K} and {H}e
  clusters formation},}\ }\href {\doibase 10.1063/1.481405} {\bibfield
  {journal} {\bibinfo  {journal} {J.\ Chem.\ Phys.}\ }\textbf {\bibinfo
  {volume} {112}},\ \bibinfo {pages} {8068--8071} (\bibinfo {year}
  {2000})}\BibitemShut {NoStop}%
\bibitem [{\citenamefont {Filsinger}\ \emph
  {et~al.}(2009{\natexlab{b}})\citenamefont {Filsinger}, \citenamefont
  {K{\"u}pper}, \citenamefont {Meijer}, \citenamefont {Hansen}, \citenamefont
  {Maurer}, \citenamefont {Nielsen}, \citenamefont {Holmegaard},\ and\
  \citenamefont {Stapelfeldt}}]{Filsinger:ACIE48:6900}%
  \BibitemOpen
  \bibfield  {author} {\bibinfo {author} {\bibfnamefont {F.}~\bibnamefont
  {Filsinger}}, \bibinfo {author} {\bibfnamefont {J.}~\bibnamefont
  {K{\"u}pper}}, \bibinfo {author} {\bibfnamefont {G.}~\bibnamefont {Meijer}},
  \bibinfo {author} {\bibfnamefont {J.~L.}\ \bibnamefont {Hansen}}, \bibinfo
  {author} {\bibfnamefont {J.}~\bibnamefont {Maurer}}, \bibinfo {author}
  {\bibfnamefont {J.~H.}\ \bibnamefont {Nielsen}}, \bibinfo {author}
  {\bibfnamefont {L.}~\bibnamefont {Holmegaard}}, \ and\ \bibinfo {author}
  {\bibfnamefont {H.}~\bibnamefont {Stapelfeldt}},\ }\bibfield  {title}
  {\enquote {\bibinfo {title} {Pure samples of individual conformers: the
  separation of stereo-isomers of complex molecules using electric fields},}\
  }\href {\doibase 10.1002/anie.200902650} {\bibfield  {journal} {\bibinfo
  {journal} {Angew.\ Chem.\ Int.\ Ed.}\ }\textbf {\bibinfo {volume} {48}},\
  \bibinfo {pages} {6900--6902} (\bibinfo {year}
  {2009}{\natexlab{b}})}\BibitemShut {NoStop}%
\bibitem [{\citenamefont {Papadakis}\ and\ \citenamefont
  {Kitsopoulos}(2006)}]{Papadakis:RSI77:3101}%
  \BibitemOpen
  \bibfield  {author} {\bibinfo {author} {\bibfnamefont {V.}~\bibnamefont
  {Papadakis}}\ and\ \bibinfo {author} {\bibfnamefont {T.~N.}\ \bibnamefont
  {Kitsopoulos}},\ }\bibfield  {title} {\enquote {\bibinfo {title} {Slice
  imaging and velocity mapping using a single field},}\ }\href {\doibase
  10.1063/1.2222084} {\bibfield  {journal} {\bibinfo  {journal} {Rev.\ Sci.\
  Instrum.}\ }\textbf {\bibinfo {volume} {77}},\ \bibinfo {pages} {3101}
  (\bibinfo {year} {2006})}\BibitemShut {NoStop}%
\bibitem [{\citenamefont {Chang}\ \emph {et~al.}(2014)\citenamefont {Chang},
  \citenamefont {Filsinger}, \citenamefont {Sartakov},\ and\ \citenamefont
  {K{\"u}pper}}]{Chang:CPC185:339}%
  \BibitemOpen
  \bibfield  {author} {\bibinfo {author} {\bibfnamefont {Y.-P.}\ \bibnamefont
  {Chang}}, \bibinfo {author} {\bibfnamefont {F.}~\bibnamefont {Filsinger}},
  \bibinfo {author} {\bibfnamefont {B.}~\bibnamefont {Sartakov}}, \ and\
  \bibinfo {author} {\bibfnamefont {J.}~\bibnamefont {K{\"u}pper}},\ }\bibfield
   {title} {\enquote {\bibinfo {title} {\textsc{CMIstark}\xspace : Python
  package for the stark-effect calculation and symmetry classification of
  linear, symmetric and asymmetric top wavefunctions in dc electric fields},}\
  }\href {\doibase 10.1016/j.cpc.2013.09.001} {\bibfield  {journal} {\bibinfo
  {journal} {Comp.\ Phys.\ Comm.}\ }\textbf {\bibinfo {volume} {185}},\
  \bibinfo {pages} {339--49} (\bibinfo {year} {2014})},\ \Eprint
  {http://arxiv.org/abs/1308.4076} {arXiv:1308.4076 [physics]} \BibitemShut
  {NoStop}%
\bibitem [{\citenamefont {Eppink}\ and\ \citenamefont
  {Parker}(1997)}]{Eppink:RSI68:3477}%
  \BibitemOpen
  \bibfield  {author} {\bibinfo {author} {\bibfnamefont {A.~T. J.~B.}\
  \bibnamefont {Eppink}}\ and\ \bibinfo {author} {\bibfnamefont {D.~H.}\
  \bibnamefont {Parker}},\ }\bibfield  {title} {\enquote {\bibinfo {title}
  {Velocity map imaging of ions and electrons using electrostatic lenses:
  Application in photoelectron and photofragment ion imaging of molecular
  oxygen},}\ }\href {\doibase 10.1063/1.1148310} {\bibfield  {journal}
  {\bibinfo  {journal} {Rev.\ Sci.\ Instrum.}\ }\textbf {\bibinfo {volume}
  {68}},\ \bibinfo {pages} {3477--3484} (\bibinfo {year} {1997})}\BibitemShut
  {NoStop}%
\bibitem [{\citenamefont {Wester}\ \emph {et~al.}(1998)\citenamefont {Wester},
  \citenamefont {Albrecht}, \citenamefont {Grieser}, \citenamefont {Knoll},
  \citenamefont {Repnow}, \citenamefont {Schwalm}, \citenamefont {Wolf},
  \citenamefont {Baer}, \citenamefont {Levin}, \citenamefont {Vager},\ and\
  \citenamefont {Zajfman}}]{Wester1998}%
  \BibitemOpen
  \bibfield  {author} {\bibinfo {author} {\bibfnamefont {R.}~\bibnamefont
  {Wester}}, \bibinfo {author} {\bibfnamefont {F.}~\bibnamefont {Albrecht}},
  \bibinfo {author} {\bibfnamefont {M.}~\bibnamefont {Grieser}}, \bibinfo
  {author} {\bibfnamefont {L.}~\bibnamefont {Knoll}}, \bibinfo {author}
  {\bibfnamefont {R.}~\bibnamefont {Repnow}}, \bibinfo {author} {\bibfnamefont
  {D.}~\bibnamefont {Schwalm}}, \bibinfo {author} {\bibfnamefont
  {A.}~\bibnamefont {Wolf}}, \bibinfo {author} {\bibfnamefont {A.}~\bibnamefont
  {Baer}}, \bibinfo {author} {\bibfnamefont {J.}~\bibnamefont {Levin}},
  \bibinfo {author} {\bibfnamefont {Z.}~\bibnamefont {Vager}}, \ and\ \bibinfo
  {author} {\bibfnamefont {D.}~\bibnamefont {Zajfman}},\ }\bibfield  {title}
  {\enquote {\bibinfo {title} {Coulomb explosion imaging at the heavy ion
  storage ring tsr},}\ }\href {\doibase 10.1016/S0168-9002(98)00553-1}
  {\bibfield  {journal} {\bibinfo  {journal} {Nuc. Instrum. \& Meth. Phys. A}\
  }\textbf {\bibinfo {volume} {413}},\ \bibinfo {pages} {379--396} (\bibinfo
  {year} {1998})}\BibitemShut {NoStop}%
\bibitem [{\citenamefont {Omiste}\ \emph {et~al.}(2011)\citenamefont {Omiste},
  \citenamefont {Gaerttner}, \citenamefont {Schmelcher}, \citenamefont
  {Gonz{\'a}lez-F{\'e}rez}, \citenamefont {Holmegaard}, \citenamefont
  {Nielsen}, \citenamefont {Stapelfeldt},\ and\ \citenamefont
  {K{\"u}pper}}]{Omiste:PCCP13:18815}%
  \BibitemOpen
  \bibfield  {author} {\bibinfo {author} {\bibfnamefont {J.~J.}\ \bibnamefont
  {Omiste}}, \bibinfo {author} {\bibfnamefont {M.}~\bibnamefont {Gaerttner}},
  \bibinfo {author} {\bibfnamefont {P.}~\bibnamefont {Schmelcher}}, \bibinfo
  {author} {\bibfnamefont {R.}~\bibnamefont {Gonz{\'a}lez-F{\'e}rez}}, \bibinfo
  {author} {\bibfnamefont {L.}~\bibnamefont {Holmegaard}}, \bibinfo {author}
  {\bibfnamefont {J.~H.}\ \bibnamefont {Nielsen}}, \bibinfo {author}
  {\bibfnamefont {H.}~\bibnamefont {Stapelfeldt}}, \ and\ \bibinfo {author}
  {\bibfnamefont {J.}~\bibnamefont {K{\"u}pper}},\ }\bibfield  {title}
  {\enquote {\bibinfo {title} {Theoretical description of adiabatic laser
  alignment and mixed-field orientation: the need for a non-adiabatic model},}\
  }\href {\doibase 10.1039/c1cp21195a} {\bibfield  {journal} {\bibinfo
  {journal} {Phys.\ Chem.\ Chem.\ Phys.}\ }\textbf {\bibinfo {volume} {13}},\
  \bibinfo {pages} {18815--18824} (\bibinfo {year} {2011})},\ \Eprint
  {http://arxiv.org/abs/1105.0534} {arXiv:1105.0534 [physics]} \BibitemShut
  {NoStop}%
\bibitem [{Note1()}]{Note1}%
  \BibitemOpen
  \bibinfo {note} {For an anti-oriented state, it holds that $-1\le \protect
  \ensuremath {\left <\protect \qopname \relax o{cos}\protect \tmspace
  -\thinmuskip {.1667em}\theta _\protect \text {2D}\right >}\protect \xspace <
  0$ and $0 \le \protect \ensuremath {\protect \text {N}_{\protect \text
  {up}}/\protect \text {N}_{\protect \text {tot}}}\protect \xspace <
  0.5$.}\BibitemShut {Stop}%
\bibitem [{Note2()}]{Note2}%
  \BibitemOpen
  \bibinfo {note} {The adiabatic pendular states \protect \ensuremath {\left
  |J,M,s\right >_\protect \textup {p}}\protect \xspace are labeled by the
  quantum numbers $J$ and $M=|m_J|$ of the adiabatically correlated field-free
  rotor states \protect \ensuremath {\left |J,M\right >}\protect \xspace and
  the symmetry $s$ with respect to reflection at the plane defined by the dc
  and ac electric field vectors, denoted $e$ for even and $o$ for odd. For the
  adiabatic following, we take into account that the dc field is turned on
  first. Once the maximum dc electric field strength is achieved, the ac laser
  pulse is turned on. These adiabatic-pendular-state labels are amended by a
  subscript $p$.}\BibitemShut {Stop}%
\bibitem [{\citenamefont {Wall}\ \emph {et~al.}(2010)\citenamefont {Wall},
  \citenamefont {Tokunaga}, \citenamefont {Hinds},\ and\ \citenamefont
  {Tarbutt}}]{Wall:PRA81:033414}%
  \BibitemOpen
  \bibfield  {author} {\bibinfo {author} {\bibfnamefont {T.~E.}\ \bibnamefont
  {Wall}}, \bibinfo {author} {\bibfnamefont {S.~K.}\ \bibnamefont {Tokunaga}},
  \bibinfo {author} {\bibfnamefont {E.~a.}\ \bibnamefont {Hinds}}, \ and\
  \bibinfo {author} {\bibfnamefont {M.~R.}\ \bibnamefont {Tarbutt}},\
  }\bibfield  {title} {\enquote {\bibinfo {title} {{Nonadiabatic transitions in
  a Stark decelerator}},}\ }\href {\doibase 10.1103/PhysRevA.81.033414}
  {\bibfield  {journal} {\bibinfo  {journal} {Phys.\ Rev.\ A}\ }\textbf
  {\bibinfo {volume} {81}},\ \bibinfo {pages} {033414} (\bibinfo {year}
  {2010})}\BibitemShut {NoStop}%
\bibitem [{\citenamefont {Strohaber}\ \emph {et~al.}(2011)\citenamefont
  {Strohaber}, \citenamefont {Mohamed}, \citenamefont {Hart}, \citenamefont
  {Zhu}, \citenamefont {Nava}, \citenamefont {Pham}, \citenamefont
  {Kolomenskii}, \citenamefont {Schroeder}, \citenamefont {Paulus},\ and\
  \citenamefont {Schuessler}}]{Strohaber:PRA84:063414}%
  \BibitemOpen
  \bibfield  {author} {\bibinfo {author} {\bibfnamefont {J.}~\bibnamefont
  {Strohaber}}, \bibinfo {author} {\bibfnamefont {T.}~\bibnamefont {Mohamed}},
  \bibinfo {author} {\bibfnamefont {N.}~\bibnamefont {Hart}}, \bibinfo {author}
  {\bibfnamefont {F.}~\bibnamefont {Zhu}}, \bibinfo {author} {\bibfnamefont
  {R.}~\bibnamefont {Nava}}, \bibinfo {author} {\bibfnamefont {F.}~\bibnamefont
  {Pham}}, \bibinfo {author} {\bibfnamefont {A.~A.}\ \bibnamefont
  {Kolomenskii}}, \bibinfo {author} {\bibfnamefont {H.}~\bibnamefont
  {Schroeder}}, \bibinfo {author} {\bibfnamefont {G.~G.}\ \bibnamefont
  {Paulus}}, \ and\ \bibinfo {author} {\bibfnamefont {H.~A.}\ \bibnamefont
  {Schuessler}},\ }\bibfield  {title} {\enquote {\bibinfo {title}
  {Intensity-resolved ionization yields of aniline with femtosecond laser
  pulses},}\ }\href {\doibase 10.1103/PhysRevA.84.063414} {\bibfield  {journal}
  {\bibinfo  {journal} {Phys.\ Rev.\ A}\ }\textbf {\bibinfo {volume} {84}},\
  \bibinfo {pages} {063414} (\bibinfo {year} {2011})}\BibitemShut {NoStop}%
\bibitem [{\citenamefont {Omiste}\ and\ \citenamefont
  {Gonz{\'a}lez-F{\'e}rez}(2013)}]{Omiste:PRA88:033416}%
  \BibitemOpen
  \bibfield  {author} {\bibinfo {author} {\bibfnamefont {J.~J.}\ \bibnamefont
  {Omiste}}\ and\ \bibinfo {author} {\bibfnamefont {R.}~\bibnamefont
  {Gonz{\'a}lez-F{\'e}rez}},\ }\bibfield  {title} {\enquote {\bibinfo {title}
  {Rotational dynamics of an asymmetric-top molecule in parallel electric and
  nonresonant laser fields},}\ }\href@noop {} {\bibfield  {journal} {\bibinfo
  {journal} {Phys.\ Rev.\ A}\ ,\ \bibinfo {pages} {033416}} (\bibinfo {year}
  {2013})}\BibitemShut {NoStop}%
\bibitem [{\citenamefont {Omiste}, \citenamefont {Gonz{\'a}lez-F{\'e}rez},\
  and\ \citenamefont {Schmelcher}(2011)}]{Omiste:JCP135:064310}%
  \BibitemOpen
  \bibfield  {author} {\bibinfo {author} {\bibfnamefont {J.~J.}\ \bibnamefont
  {Omiste}}, \bibinfo {author} {\bibfnamefont {R.}~\bibnamefont
  {Gonz{\'a}lez-F{\'e}rez}}, \ and\ \bibinfo {author} {\bibfnamefont
  {P.}~\bibnamefont {Schmelcher}},\ }\bibfield  {title} {\enquote {\bibinfo
  {title} {{Rotational spectrum of asymmetric top molecules in combined static
  and laser fields}},}\ }\href {\doibase 10.1063/1.3624774} {\bibfield
  {journal} {\bibinfo  {journal} {J.\ Chem.\ Phys.}\ }\textbf {\bibinfo
  {volume} {135}},\ \bibinfo {pages} {064310} (\bibinfo {year} {2011})},\
  \Eprint {http://arxiv.org/abs/1106.1586} {arXiv:1106.1586 [physics.chem-ph]}
  \BibitemShut {NoStop}%
\bibitem [{\citenamefont {Filsinger}\ \emph {et~al.}(2008)\citenamefont
  {Filsinger}, \citenamefont {Erlekam}, \citenamefont {von Helden},
  \citenamefont {K{\"u}pper},\ and\ \citenamefont
  {Meijer}}]{Filsinger:PRL100:133003}%
  \BibitemOpen
  \bibfield  {author} {\bibinfo {author} {\bibfnamefont {F.}~\bibnamefont
  {Filsinger}}, \bibinfo {author} {\bibfnamefont {U.}~\bibnamefont {Erlekam}},
  \bibinfo {author} {\bibfnamefont {G.}~\bibnamefont {von Helden}}, \bibinfo
  {author} {\bibfnamefont {J.}~\bibnamefont {K{\"u}pper}}, \ and\ \bibinfo
  {author} {\bibfnamefont {G.}~\bibnamefont {Meijer}},\ }\bibfield  {title}
  {\enquote {\bibinfo {title} {Selector for structural isomers of neutral
  molecules},}\ }\href {\doibase 10.1103/PhysRevLett.100.133003} {\bibfield
  {journal} {\bibinfo  {journal} {Phys.\ Rev.\ Lett.}\ }\textbf {\bibinfo
  {volume} {100}},\ \bibinfo {pages} {133003} (\bibinfo {year} {2008})},\
  \Eprint {http://arxiv.org/abs/0802.2795} {arXiv:0802.2795 [physics]}
  \BibitemShut {NoStop}%
\bibitem [{\citenamefont {Filsinger}\ \emph {et~al.}(2010)\citenamefont
  {Filsinger}, \citenamefont {Putzke}, \citenamefont {Haak}, \citenamefont
  {Meijer},\ and\ \citenamefont {K{\"u}pper}}]{Filsinger:PRA82:052513}%
  \BibitemOpen
  \bibfield  {author} {\bibinfo {author} {\bibfnamefont {F.}~\bibnamefont
  {Filsinger}}, \bibinfo {author} {\bibfnamefont {S.}~\bibnamefont {Putzke}},
  \bibinfo {author} {\bibfnamefont {H.}~\bibnamefont {Haak}}, \bibinfo {author}
  {\bibfnamefont {G.}~\bibnamefont {Meijer}}, \ and\ \bibinfo {author}
  {\bibfnamefont {J.}~\bibnamefont {K{\"u}pper}},\ }\bibfield  {title}
  {\enquote {\bibinfo {title} {Optimizing the resolution of the
  alternating-gradient $m/\mu$ selector},}\ }\href {\doibase
  10.1103/PhysRevA.82.052513} {\bibfield  {journal} {\bibinfo  {journal}
  {Phys.\ Rev.\ A}\ }\textbf {\bibinfo {volume} {82}},\ \bibinfo {pages}
  {052513} (\bibinfo {year} {2010})}\BibitemShut {NoStop}%
\bibitem [{\citenamefont {Pullen}\ \emph {et~al.}(2015)\citenamefont {Pullen},
  \citenamefont {Wolter}, \citenamefont {Le}, \citenamefont {Baudisch},
  \citenamefont {Hemmer}, \citenamefont {Senftleben}, \citenamefont {Schroter},
  \citenamefont {Ullrich}, \citenamefont {Moshammer}, \citenamefont {Lin},\
  and\ \citenamefont {Biegert}}]{Pullen:NatComm6:7262}%
  \BibitemOpen
  \bibfield  {author} {\bibinfo {author} {\bibfnamefont {M.~G.}\ \bibnamefont
  {Pullen}}, \bibinfo {author} {\bibfnamefont {B.}~\bibnamefont {Wolter}},
  \bibinfo {author} {\bibfnamefont {A.-T.}\ \bibnamefont {Le}}, \bibinfo
  {author} {\bibfnamefont {M.}~\bibnamefont {Baudisch}}, \bibinfo {author}
  {\bibfnamefont {M.}~\bibnamefont {Hemmer}}, \bibinfo {author} {\bibfnamefont
  {A.}~\bibnamefont {Senftleben}}, \bibinfo {author} {\bibfnamefont {C.~D.}\
  \bibnamefont {Schroter}}, \bibinfo {author} {\bibfnamefont {J.}~\bibnamefont
  {Ullrich}}, \bibinfo {author} {\bibfnamefont {R.}~\bibnamefont {Moshammer}},
  \bibinfo {author} {\bibfnamefont {C.~D.}\ \bibnamefont {Lin}}, \ and\
  \bibinfo {author} {\bibfnamefont {J.}~\bibnamefont {Biegert}},\ }\bibfield
  {title} {\enquote {\bibinfo {title} {Imaging an aligned polyatomic molecule
  with laser-induced electron diffraction},}\ }\href {\doibase
  10.1038/ncomms8262} {\bibfield  {journal} {\bibinfo  {journal} {Nat.
  Commun.}\ }\textbf {\bibinfo {volume} {6}},\ \bibinfo {pages} {7262}
  (\bibinfo {year} {2015})}\BibitemShut {NoStop}%
\bibitem [{\citenamefont {Zeidler}\ \emph {et~al.}(2005)\citenamefont
  {Zeidler}, \citenamefont {Staudte}, \citenamefont {Bardon}, \citenamefont
  {Villeneuve}, \citenamefont {D{\"o}rner},\ and\ \citenamefont
  {Corkum}}]{Zeidler:PRL95:203003}%
  \BibitemOpen
  \bibfield  {author} {\bibinfo {author} {\bibfnamefont {D.}~\bibnamefont
  {Zeidler}}, \bibinfo {author} {\bibfnamefont {A.}~\bibnamefont {Staudte}},
  \bibinfo {author} {\bibfnamefont {A.~B.}\ \bibnamefont {Bardon}}, \bibinfo
  {author} {\bibfnamefont {D.~M.}\ \bibnamefont {Villeneuve}}, \bibinfo
  {author} {\bibfnamefont {R.}~\bibnamefont {D{\"o}rner}}, \ and\ \bibinfo
  {author} {\bibfnamefont {P.~B.}\ \bibnamefont {Corkum}},\ }\bibfield  {title}
  {\enquote {\bibinfo {title} {Controlling attosecond double ionization
  dynamics via molecular alignment},}\ }\href
  {http://link.aps.org/abstract/PRL/v95/e203003} {\bibfield  {journal}
  {\bibinfo  {journal} {Phys.\ Rev.\ Lett.}\ }\textbf {\bibinfo {volume}
  {95}},\ \bibinfo {eid} {203003} (\bibinfo {year} {2005})}\BibitemShut
  {NoStop}%
\bibitem [{\citenamefont {Boll}\ \emph {et~al.}(2014)\citenamefont {Boll},
  \citenamefont {Rouz{\'e}e}, \citenamefont {Adolph}, \citenamefont {Anielski},
  \citenamefont {Aquila}, \citenamefont {Bari}, \citenamefont {Bomme},
  \citenamefont {Bostedt}, \citenamefont {Bozek}, \citenamefont {Chapman},
  \citenamefont {Christensen}, \citenamefont {Coffee}, \citenamefont {Coppola},
  \citenamefont {De}, \citenamefont {Decleva}, \citenamefont {Epp},
  \citenamefont {Erk}, \citenamefont {Filsinger}, \citenamefont {Foucar},
  \citenamefont {Gorkhover}, \citenamefont {Gumprecht}, \citenamefont
  {H{\"o}mke}, \citenamefont {Holmegaard}, \citenamefont {Johnsson},
  \citenamefont {Kienitz}, \citenamefont {Kierspel}, \citenamefont {Krasniqi},
  \citenamefont {K{\"u}hnel}, \citenamefont {Maurer}, \citenamefont
  {Messerschmidt}, \citenamefont {Moshammer}, \citenamefont {M{\"u}ller},
  \citenamefont {Rudek}, \citenamefont {Savelyev}, \citenamefont {Schlichting},
  \citenamefont {Schmidt}, \citenamefont {Scholz}, \citenamefont {Schorb},
  \citenamefont {Schulz}, \citenamefont {Seltmann}, \citenamefont {Stener},
  \citenamefont {Stern}, \citenamefont {Techert}, \citenamefont {Th{\o}gersen},
  \citenamefont {Trippel}, \citenamefont {Viefhaus}, \citenamefont {Vrakking},
  \citenamefont {Stapelfeldt}, \citenamefont {K{\"u}pper}, \citenamefont
  {Ullrich}, \citenamefont {Rudenko},\ and\ \citenamefont
  {Rolles}}]{Boll:FD17171:57}%
  \BibitemOpen
  \bibfield  {author} {\bibinfo {author} {\bibfnamefont {R.}~\bibnamefont
  {Boll}}, \bibinfo {author} {\bibfnamefont {A.}~\bibnamefont {Rouz{\'e}e}},
  \bibinfo {author} {\bibfnamefont {M.}~\bibnamefont {Adolph}}, \bibinfo
  {author} {\bibfnamefont {D.}~\bibnamefont {Anielski}}, \bibinfo {author}
  {\bibfnamefont {A.}~\bibnamefont {Aquila}}, \bibinfo {author} {\bibfnamefont
  {S.}~\bibnamefont {Bari}}, \bibinfo {author} {\bibfnamefont {C.}~\bibnamefont
  {Bomme}}, \bibinfo {author} {\bibfnamefont {C.}~\bibnamefont {Bostedt}},
  \bibinfo {author} {\bibfnamefont {J.~D.}\ \bibnamefont {Bozek}}, \bibinfo
  {author} {\bibfnamefont {H.~N.}\ \bibnamefont {Chapman}}, \bibinfo {author}
  {\bibfnamefont {L.}~\bibnamefont {Christensen}}, \bibinfo {author}
  {\bibfnamefont {R.}~\bibnamefont {Coffee}}, \bibinfo {author} {\bibfnamefont
  {N.}~\bibnamefont {Coppola}}, \bibinfo {author} {\bibfnamefont
  {S.}~\bibnamefont {De}}, \bibinfo {author} {\bibfnamefont {P.}~\bibnamefont
  {Decleva}}, \bibinfo {author} {\bibfnamefont {S.~W.}\ \bibnamefont {Epp}},
  \bibinfo {author} {\bibfnamefont {B.}~\bibnamefont {Erk}}, \bibinfo {author}
  {\bibfnamefont {F.}~\bibnamefont {Filsinger}}, \bibinfo {author}
  {\bibfnamefont {L.}~\bibnamefont {Foucar}}, \bibinfo {author} {\bibfnamefont
  {T.}~\bibnamefont {Gorkhover}}, \bibinfo {author} {\bibfnamefont
  {L.}~\bibnamefont {Gumprecht}}, \bibinfo {author} {\bibfnamefont
  {A.}~\bibnamefont {H{\"o}mke}}, \bibinfo {author} {\bibfnamefont
  {L.}~\bibnamefont {Holmegaard}}, \bibinfo {author} {\bibfnamefont
  {P.}~\bibnamefont {Johnsson}}, \bibinfo {author} {\bibfnamefont {J.~S.}\
  \bibnamefont {Kienitz}}, \bibinfo {author} {\bibfnamefont {T.}~\bibnamefont
  {Kierspel}}, \bibinfo {author} {\bibfnamefont {F.}~\bibnamefont {Krasniqi}},
  \bibinfo {author} {\bibfnamefont {K.-U.}\ \bibnamefont {K{\"u}hnel}},
  \bibinfo {author} {\bibfnamefont {J.}~\bibnamefont {Maurer}}, \bibinfo
  {author} {\bibfnamefont {M.}~\bibnamefont {Messerschmidt}}, \bibinfo {author}
  {\bibfnamefont {R.}~\bibnamefont {Moshammer}}, \bibinfo {author}
  {\bibfnamefont {N.~L.~M.}\ \bibnamefont {M{\"u}ller}}, \bibinfo {author}
  {\bibfnamefont {B.}~\bibnamefont {Rudek}}, \bibinfo {author} {\bibfnamefont
  {E.}~\bibnamefont {Savelyev}}, \bibinfo {author} {\bibfnamefont
  {I.}~\bibnamefont {Schlichting}}, \bibinfo {author} {\bibfnamefont
  {C.}~\bibnamefont {Schmidt}}, \bibinfo {author} {\bibfnamefont
  {F.}~\bibnamefont {Scholz}}, \bibinfo {author} {\bibfnamefont
  {S.}~\bibnamefont {Schorb}}, \bibinfo {author} {\bibfnamefont
  {J.}~\bibnamefont {Schulz}}, \bibinfo {author} {\bibfnamefont
  {J.}~\bibnamefont {Seltmann}}, \bibinfo {author} {\bibfnamefont
  {M.}~\bibnamefont {Stener}}, \bibinfo {author} {\bibfnamefont
  {S.}~\bibnamefont {Stern}}, \bibinfo {author} {\bibfnamefont
  {S.}~\bibnamefont {Techert}}, \bibinfo {author} {\bibfnamefont
  {J.}~\bibnamefont {Th{\o}gersen}}, \bibinfo {author} {\bibfnamefont
  {S.}~\bibnamefont {Trippel}}, \bibinfo {author} {\bibfnamefont
  {J.}~\bibnamefont {Viefhaus}}, \bibinfo {author} {\bibfnamefont
  {M.}~\bibnamefont {Vrakking}}, \bibinfo {author} {\bibfnamefont
  {H.}~\bibnamefont {Stapelfeldt}}, \bibinfo {author} {\bibfnamefont
  {J.}~\bibnamefont {K{\"u}pper}}, \bibinfo {author} {\bibfnamefont
  {J.}~\bibnamefont {Ullrich}}, \bibinfo {author} {\bibfnamefont
  {A.}~\bibnamefont {Rudenko}}, \ and\ \bibinfo {author} {\bibfnamefont
  {D.}~\bibnamefont {Rolles}},\ }\bibfield  {title} {\enquote {\bibinfo {title}
  {Imaging molecular structure through femtosecond photoelectron diffraction on
  aligned and oriented gas-phase molecules},}\ }\href {\doibase
  10.1039/c4fd00037d} {\bibfield  {journal} {\bibinfo  {journal} {Faraday
  Disc.}\ }\textbf {\bibinfo {volume} {171}},\ \bibinfo {pages} {57–80}
  (\bibinfo {year} {2014})},\ \Eprint {http://arxiv.org/abs/1407.7782}
  {arXiv:1407.7782} \BibitemShut {NoStop}%
\end{thebibliography}%
\end{document}